\documentclass[reprint,aps,prb,floatfix,twocolumn,superscriptaddress,showpacs,amsmath,amssymb]{revtex4-1}
\usepackage{color,graphicx,calc,url,float}
\usepackage{dcolumn}
\usepackage{bm}
\usepackage{amssymb}
\usepackage{amsmath}
\usepackage{wasysym}
\usepackage[colorlinks, 
			linkcolor=blue,
			citecolor=blue,
			urlcolor=blue
			]{hyperref}
\usepackage{tabularx}
\usepackage{longtable}

\graphicspath{{../figures/}}

\begin{document}
\title{Enhanced magnetization of ultrathin NiFe$_\text{2}$O$_\text{4}$ films on SrTiO$_\text{3}$(001) related to cation disorder and anomalous strain}
\author{J. Rodewald}
\affiliation{Department of Physics, Osnabr\"uck University, Barbarastra\ss e 7, D-49076 Osnabr\"uck, Germany}
\author{J. Thien}
\affiliation{Department of Physics, Osnabr\"uck University, Barbarastra\ss e 7, D-49076 Osnabr\"uck, Germany}
\author{K. Ruwisch}
\affiliation{Department of Physics, Osnabr\"uck University, Barbarastra\ss e 7, D-49076 Osnabr\"uck, Germany}
\author{F. Bertram}
\affiliation{DESY Photon Science, Notkestra\ss e 85, D-22607 Hamburg, Germany}
\author{K. Kuepper}
\affiliation{Department of Physics, Osnabr\"uck University, Barbarastra\ss e 7, D-49076 Osnabr\"uck, Germany}
\author{J. Wollschl\"ager}
\affiliation{Department of Physics, Osnabr\"uck University, Barbarastra\ss e 7, D-49076 Osnabr\"uck, Germany}

\date{\today}


\begin{abstract}
NiFe$_2$O$_4$ thin films with varying thickness were grown on SrTiO$_3$(001) by reactive molecular beam epitaxy. 
Soft and hard x-ray photoelectron spectroscopy measurements reveal a homogeneous cation distribution throughout the whole film with stoichiometric Ni:Fe ratios of 1:2 independent of the film thickness. Low energy electron diffraction and high resolution (grazing incidence) x-ray diffraction in addition to x-ray reflectivity experiments were conducted to obtain information of the film surface and bulk structure, respectively. For ultrathin films up to 7.3 nm, lateral tensile and vertical compressive strain is observed, contradicting an adaption at the interface of NiFe$_2$O$_4$ film and substrate lattice. The applied strain is accompanied by an increased lateral defect density, which is decaying for relaxed thicker films and attributed to the growth of lateral grains. Determination of cationic site occupancies in the inverse spinel structure by analysis of site sensitive diffraction peaks reveals low tetrahedral occupancies for thin, strained NiFe$_2$O$_4$ films, resulting in partial presence of deficient rock salt like structures. These structures are assumed to be responsible for the enhanced magnetization of up to $\sim$250\% of the NiFe$_2$O$_4$ bulk magnetization as observed by superconducting quantum interference device magnetometry for ultrathin films below 7.3 nm thickness.
\end{abstract}

\maketitle

\section{Introduction}
Complex oxides are in the focus of current research due to the competitive interaction of spin, charge and orbital degrees of freedom, resulting in a variety of intriguing phenomena like ferromagnetism, ferroelectricity or multiferroicity \cite{Hwang2012}. 
In particular, transition metal ferrites are of great interest in the fields of spin-based electronics, e.g., spincaloritronics \cite{Bauer2012} and spintronics \cite{Hoffmann2015, Cibert2005, Moussy2013}, as the requirements of high Curie temperatures T$_\text{C}$ and significant magnetic saturation moments are fulfilled \cite{Brabers1995}.

In the field of spintronics, the quality of spin-based devices is determined by the capability of generating highly spin-polarized electron currents. 
Here, NiFe$_2$O$_4$ (NFO) as an insulating and ferrimagnetic inverse spinel up to a high T$_\text{C}$ of 865 K \cite{Brabers1995} represents a promising candidate for spin-filter applications \cite{Lueders2006, Matzen2014}. 
As the spin-tunneling probability decreases exponentially with increasing tunneling barrier thickness, ferrites have to be prepared as ultrathin films with low roughness if high-quality spin filters are aimed for.

As a cubic inverse spinel, NFO exhibits a rather complex structure: O$^{2-}$ anions form an fcc (face-centered cubic) sublattice, whereas Ni$^{2+}$ cations occupy octahedrally ($O_h$) coordinated lattice sites and Fe$^{3+}$ cations are equally distributed across both $O_h$ and tetrahedral ($T_d$) sites. 
The distribution of cations affects most properties of the ferrites. 
In particular, magnetic ordering is dominated by oxygen mediated superexchange interaction between $T_d$ and $O_h$ coordinated cations on two antiferromagnetically coupled sublattices in addition to oxygen mediated double-exchange interaction of Fe$^{3+}$ and Ni$^{2+}$ cations on $O_h$ sites with ferromagnetic coupling.

In most studies, thin ferrite films are deposited on MgO(001) substrates as the lattice mismatch between, i.e., NFO (lattice constant $a_\text{NFO}$=833.9 pm) and MgO(001) ($a_\text{MgO}$=421.2 pm) is only 1\% (comparing the halved lattice constant of NFO with one lattice distance of MgO). 
Moreover, induced strain in the film material can have significant impact on the interplay between its spin, charge and orbital degrees of freedom.  
This aspect can be used to tune the physical properties of, e.g., ferrite films by applying epitaxially induced strain during the controlled growth on substrates with a larger lattice mismatch. 
Therefore, SrTiO$_3$(001) (STO) with a lattice mismatch of 6.8\% ($a_{\text{STO}}$=390.5 pm) compared to NFO(001) is used as a substrate in this work. 
Here, strain applied at the interface between substrate and film may alter the physical properties of the NFO film. 
Thus, this work focuses on the influence of substrate induced strain applied at the interface on structural and magnetic properties of thin NFO films.

Various deposition techniques are used to prepare thin ferrite films as, e.g., sputter deposition \cite{Lueders2006, Klewe2014}, pulsed laser deposition \cite{Hoppe2015} or chemical solution deposition \cite{Seifikar2012}. 
Here, we grow thin NFO films on STO(001) by using reactive molecular beam epitaxy (RMBE). 
For potential spin-based applications, knowledge about fundamental chemical and structural, as well as magnetic properties is indispensable.

For this purpose, NFO thin films of different film thicknesses have been prepared. 
The film surface structure and composition has been studied by means of low energy electron diffraction (LEED) and x-ray photoelectron spectroscopy (XPS), respectively. 
Bulk composition and structure are analyzed by hard x-ray photoelectron spectroscopy (HAXPES) and high resolution x-ray diffraction (HR-XRD) experiments, respectively. 
In addition, film thicknesses and roughnesses are determined by x-ray reflectivity (XRR) measurements. 
The magnetic properties of the films are characterized using superconducting quantum interference device (SQUID) magnetometry.

\section{Experimental Details}
\label{sec:Experimental Details}
NFO films of different thicknesses between 3.7 and 55.5 nm have been prepared on conductive 0.05 wt\% Nb doped STO(001) substrates via RMBE in an ultra high vacuum (UHV) chamber. 
Prior to film deposition, the substrates have been cleaned by annealing at 400$^\circ$C for 1h in a molecular oxygen atmosphere of 1$\times$10$^{-4}$ mbar. 
The thin ferrite films have been deposited by thermal co-evaporation of the metals from pure rods in a diluted molecular oxygen atmosphere of 5$\times$10$^{-6}$ mbar. 
The substrate temperature was held at 250$^\circ$C and the deposition rate was controlled to 1 nm/min for all films. 
Chemical composition and crystallinity of the substrate and film surface were characterized \textit{in situ} by means of XPS and LEED, respectively. 
For XPS measurements, Mg K$\alpha$ radiation with a photon energy of E$_\text{ph}$ = 1253.6 eV was used.

In order to gain insight into the chemical composition and oxidation states in deeper subsurface layers HAXPES measurements were performed at beamline P22 of the Deutsches Elektronen-Synchrotron (DESY), Germany. 
HAXPES spectra were recorded using a SPECS Phoibos 225 HV hemispherical analyzer with a delay-line detector. 
Here, a photon energy of E$_\text{ph}$ = 6 keV was used to create photoelectrons with high kinetic energy and thus, a higher probing depth than for the lab-based \textit{in situ} XPS measurements.

Complementary to XPS and HAXPES, bulk sensitive x-ray fluorescence (XRF) spectroscopy measurements were performed, using an in-house-built XRF setup at Bielefeld University, Germany.

Besides, XRR measurements were conducted in a Philips X'Pert Pro diffractometer at Bielefeld University to determine film thicknesses and surface roughnesses, using Cu K${\alpha}$ radiation with a photon energy of  E$_\text{ph}(\text{Cu}_{K\alpha})$ = 8048.0 eV. 
XRR data was analyzed, using an in-house-developed fitting tool based on the Parratt algorithm \cite{Parratt1954} and N\'{e}vot-Croce roughness model \cite{Nevot1980}.

In addition, HR-XRD experiments were performed in order to obtain structural information of the deposited NFO films. 
Therefore, scans along the (00L) crystal truncation rod (CTR) were performed in $\theta-2\theta$ geometry close to the STO(002) Bragg reflection. 
Additional x-ray diffraction measurements along (H00) direction and along the NFO(22L) CTR were performed in grazing incidence geometry (GIXRD) with a fixed incident angle of 0.3$^\circ$ between the incoming x-ray beam and the sample surface. 
For these (GI)XRD measurements, the samples were transported to beamline P08 of DESY, where a photon energy of E$_\text{ph}$ = 18 keV and a six-circle diffractometer with a two-dimensional PILATUS 100k detector was used. 

Magnetic properties were characterized, using a SQUID of type S700X from CRYOGENIC.

\section{Results}
\subsection{Surface characterization: LEED and XPS}
\label{subsec:LEEDandXPS}
\begin{figure*}[t]
	\centering
	\includegraphics[width=\textwidth]{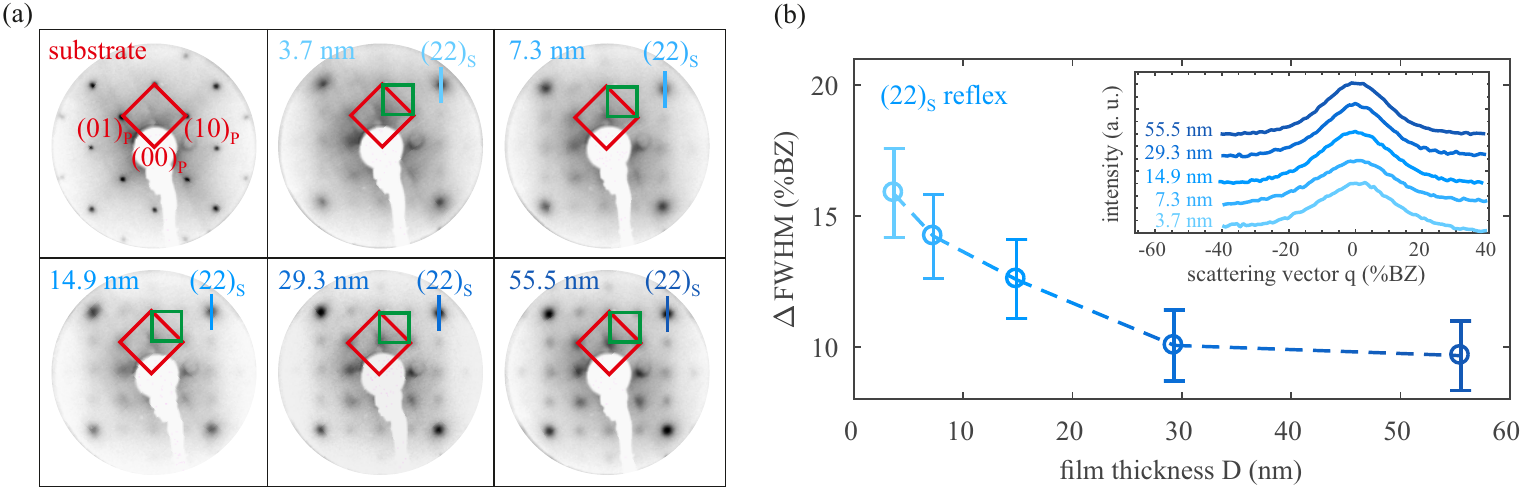}
	\caption{(a) LEED images for the prepared substrate and NFO films at an electron energy of 98 eV. The square (1$\times$1) reciprocal surface unit cell of the spinel type NFO films (green square) is half the size and rotated by 45$^\circ$ with respect to the (1$\times$1) surface structure of the perovskite type STO(001) substrate (red square). (b) $\Delta$FWHM given as the FWHM of the (22)$_\text{S}$ reflex for the different film thicknesses corrected by the instrumental broadening, which is estimated by the FWHM of the (20)$_\text{P}$ reflex of the substrate. The unit of $\Delta$FWHM is given in \% of the first Brillouin zone of STO(001) (100\%BZ = 2$\pi$/a$_\text{STO}$). The (22)$_\text{S}$ spot profiles (shown in the inset) were extracted from the 2D LEED images [marked in (a) by a blue line].}
	\label{fig:LEED}
\end{figure*}

A structural characterization of NFO film surfaces was examined \textit{in situ} directly after film deposition by means of LEED. 
The two dimensional LEED pattern of the STO substrate and the NFO films of different thicknesses at an electron energy of 98 eV are depicted in Fig. \ref{fig:LEED} (a). 
The diffraction pattern of the cleaned substrate clearly exhibits a square (1$\times$1) surface structure with sharp and intense diffraction spots, stemming from a well-ordered crystalline STO(001) surface. 
Even for the thinnest NFO layer with a thickness of 3.7 nm this pattern vanishes due to the surface sensitivity of LEED. 
Instead, only weak and broad diffraction spots close to the positions of the (20)$_\text{P}$, (02)$_\text{P}$, ($\bar{2}$0)$_\text{P}$ and (0$\bar{2}$)$_\text{P}$ spots of the substrate pattern get visible. 
This indicates a fully closed NFO film covering the entire STO surface. 
Here, the LEED spots are denoted by indices P and S to distinguish between perovskite type (STO) and spinel type (NFO) surface unit cells, respectively. 
The low intensity to background ratio in the LEED pattern of the thinnest film points to a rough NFO film surface with many defects. 
For further comparison, the (1$\times$1) substrate surface structure is marked by a red square in every LEED pattern of the NFO films in Fig. \ref{fig:LEED} (a), although its diffraction spots are not visible anymore. 
For the film of 7.3 nm thickness, the NFO film reflexes rise in intensity and further spots emerge, marking a square (1$\times$1) surface structure (green square). 
This reciprocal surface unit cell is rotated by 45$^{\circ}$ with respect to the surface unit cell of STO with half the unit cell size, as it is expected for the growth of spinel type NFO(001) with an fcc oxygen sublattice on the perovskite STO(001). 
Thus, the $\lbrace$22$\rbrace$$_\text{S}$ diffraction spots almost coincide with the $\lbrace$20$\rbrace$$_\text{P}$ diffraction spots, pointing to cube-on-cube epitaxy. 
The NFO surface diffraction spots get even more pronounced for thicker films up to 55.5 nm, pointing to less defects and smoother surfaces for thicker NFO films. 
This behavior comes along with a decreasing full width at half maximum (FWHM) of the reflexes, exemplarily shown in Fig. \ref{fig:LEED} (b) for the (22)$_\text{S}$ diffraction spot. 
This reflex is also marked by a blue line in the 2D pattern in Fig. \ref{fig:LEED} (a) and solely originates from diffraction of a crystalline NFO(001) surface. 
The FWHM is given in \% of the first Brillouin zone of the STO(001) substrate, which means that 100\%BZ = 2$\pi$/a$_\text{STO}$.
The FWHM of the sharp (20)$_\text{P}$ reflex of the substrate serves as a measure for the instrumental broadening of the diffraction spots. 
Thus, $\Delta$FWHM is given by the difference between the FWHM of the (22)$_\text{S}$ and the (20)$_\text{P}$ reflex and is therefore solely determined by the average defect-defect distance (domain or terrace size). 
Here, the decreasing $\Delta$FWHM points to an increasing long-range order of the NFO film surface for thicker films, reaching a minimum value of $\Delta$FWHM$\approx$10\%BZ for the thickest film, meaning a domain size of about ten unit cells of STO(001).

\begin{figure}[b]
	\centering
	\includegraphics[width=\columnwidth]{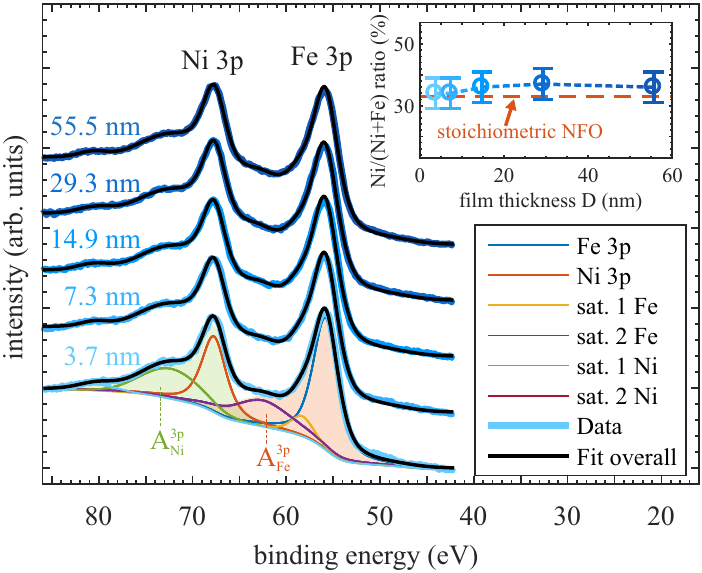}
	\caption{Ni 3p and Fe 3p spectra measured by XPS for all prepared NFO films with varying thickness. Both Ni 3p and Fe 3p spectra consist of the main 3p signal and two satellites at their higher binding energy sides. The resulting cationic Ni amount (shown in the inset) is constant for all film thicknesses and considering an experimental error of 5\% includes the stoichiometric value of 33\%.}
	\label{fig:XPS_Ni3pFe3p}
\end{figure}

Chemical composition of the NFO films was determined \textit{in situ} by surface sensitive XPS. 
Therefore, Ni 3p and Fe 3p core level spectra, shown in Fig. \ref{fig:XPS_Ni3pFe3p}, were fitted by Voight functions after subtracting a Shirley background. 
Each Ni 3p and Fe 3p spectrum consists of a main 3p peak and two satellites at higher binding energies, which contribute all to the whole intensity $I$ of the respective spectra. 
All contributions with the related overall fit are exemplarily depicted in Fig. \ref{fig:XPS_Ni3pFe3p} for the thinnest NFO film of 3.7 nm. 
As binding energies of Ni 3p and Fe 3p spectra do not differ significantly from each other, the energy dependent inelastic mean free path of the photoelectrons and the transmission function of the spectrometer are similar as well. 
This leads to an intensity yield and thereby to a cationic Ni amount of
\begin{equation}
	\label{eq:IntRatioXPS}
	Y_{Ni} = \frac{I_{Ni}}{I_{Ni} + I_{Fe}} = \frac{A_{Ni}^{3p}/\sigma_{Ni}^{3p}}{A_{Ni}^{3p}/\sigma_{Ni}^{3p} + A_{Fe}^{3p}/\sigma_{Fe}^{3p}}
\end{equation}
with $A_{Ni,Fe}^{3p}$ as the background corrected area below the 3p spectra and $\sigma_{Ni,Fe}^{3p}$ as the respective photoelectric cross sections \cite{Scofield1976}. 
For all prepared films, similar shapes and intensities of Fe and Ni 3p core level spectra were observed.
In fact, quantitative analysis of the cationic Ni amount by applying equation (\ref{eq:IntRatioXPS}) results in values between 34\% and 37\% (cf. Table \ref{tab:compositions}), stating that all NFO films exhibit a similar Ni amount and - considering an experimental error of $\pm$5\% - include the stoichiometric value of 33\%.

\begin{figure*}[]
	\centering
	\begin{minipage}[c]{\columnwidth}
		\includegraphics[width=1\columnwidth]{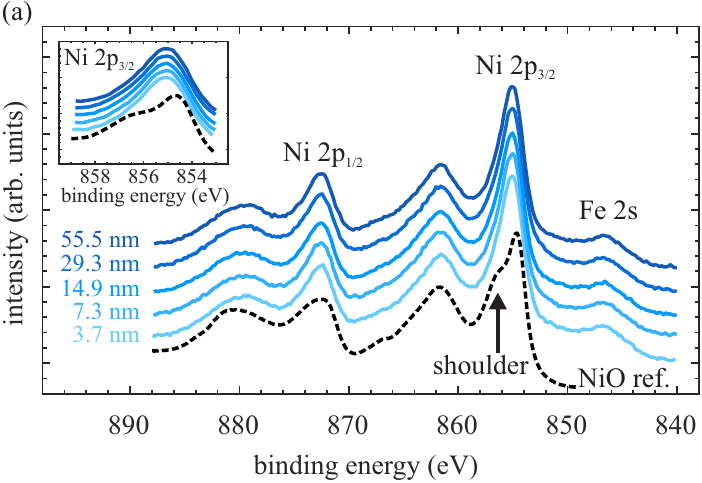}
		\label{fig:XPS_Ni2p}
	\end{minipage}
	\begin{minipage}[c]{\columnwidth}
		\includegraphics[width=1\columnwidth]{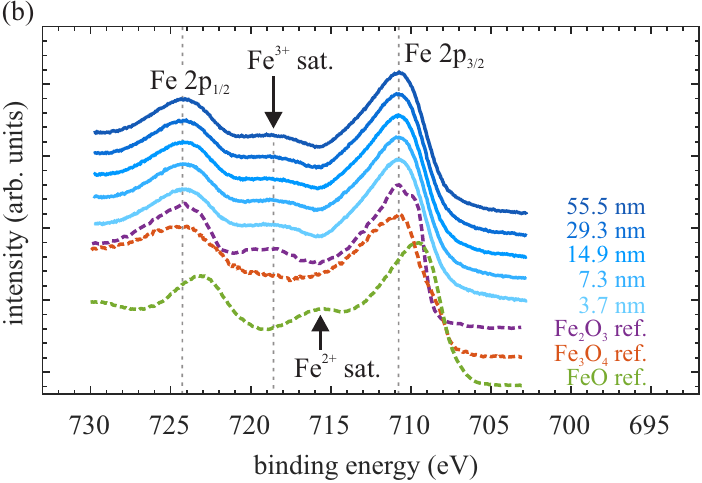}
		\label{fig:XPS_Fe2p}
	\end{minipage}
	\caption{XPS measurements of all prepared NFO films with varying film thickness. (a) Ni 2p core level spectra and a reference spectrum of a 12 nm thin NiO film. The Ni 2p$_{3/2}$ signals are shown in the inset on a smaller scale for better comparison. (b) Fe 2p core level spectra compared to Fe$_2$O$_3$, Fe$_3$O$_4$ and FeO reference spectra.}
	\label{fig:XPS_Ni2pFe2p}
\end{figure*}
Additionally, Ni 2p and Fe 2p spectra (cf. Fig. \ref{fig:XPS_Ni2pFe2p}) are used to qualitatively estimate the major valence state of Ni and Fe cations in the near-surface region of the prepared NFO films, respectively. 
The Ni 2p spectra in Fig. \ref{fig:XPS_Ni2pFe2p} (a) show no significant difference for all film thicknesses. 
The spectra consist of two main spin-orbit split Ni 2p$_{1/2}$ and Ni 2p$_{3/2}$ peaks accompanied by satellite peaks $\sim$7 eV above the binding energies of the main peaks. 
In addition, the Ni 2p spectra are superimposed by the Fe 2s signal at the low binding energy side of the Ni 2p$_{3/2}$ signal. 
The position of the main Ni 2p$_{3/2}$ peaks with a binding energy of 855.1(2) eV remains constant for all films and is consistent with reported values for stoichiometric NFO \cite{McIntyre1975}. 
For comparison, a Ni 2p reference spectrum of a 12 nm thin NiO film on STO(001) measured by the same XPS system is shown additionally in Fig. \ref{fig:XPS_Ni2pFe2p} (a). 
The position of the Ni 2p$_{3/2}$ peak in NiO with a binding energy of 854.6 eV clearly differs from the measured value of the NFO films [see inset of Fig. \ref{fig:XPS_Ni2pFe2p} (a)]. 
Furthermore, there is no shoulder on the high binding energy side of the Ni 2p$_{3/2}$ peak, as it is observed in the NiO reference spectrum. 
This shoulder is theoretically described as a result of a screening process by electrons not originating from the oxygen orbitals around the Ni atom, but from neighbouring NiO$_6$ clusters and is therefore characteristic for NiO films thicker than one monolayer \cite{Veenendaal1993, Alders1996}. 
Thus, the presence of NiO clusters in the near-surface region of the prepared NFO films can be excluded.

As seen in Fig. \ref{fig:XPS_Ni2pFe2p} (b), the respective Fe 2p spectra do also not differ significantly in shape and intensity from each other. 
The Fe 2p$_{3/2}$ and Fe 2p$_{1/2}$ positions with binding energies of 710.8(2) eV and 724.3(3) eV, respectively, coincide with reported values of Fe$_3$O$_4$ and Fe$_2$O$_3$ \cite{Yamashita2008}. 
In addition, a satellite between Fe 2p$_{3/2}$ and Fe 2p$_{1/2}$ at 718.7(3) eV is observed, corresponding to the satellite in Fe$_2$O$_3$ [see Fe$_2$O$_3$ reference spectrum in Fig. \ref{fig:XPS_Ni2pFe2p} (b) for comparison] \cite{Yamashita2008}. 
In contrast, Fe$_3$O$_4$ spectra do not exhibit a distinct satellite between the Fe 2p$_{3/2}$ and Fe 2p$_{1/2}$ signals due to the overlap of satellites resulting from Fe$^{2+}$ (cf. satellite in FeO reference spectrum) and Fe$^{3+}$ cations. 
Therefore, the observed shape of the Fe 2p spectra of the prepared NFO films can be ascribed to Fe$^{3+}$ excess. 
These observations made in XPS lead to the conclusion, that all NFO films exhibit stoichiometric Ni/Fe ratios with an excess of Fe$^{3+}$ cations and negligible Fe$^{2+}$ contributions, as it is expected for stoichiometric NFO films.

\subsection{HAXPES}
\label{subsec:HAXPES}

\begin{figure}[b]
	\centering
	\includegraphics[width=\columnwidth]{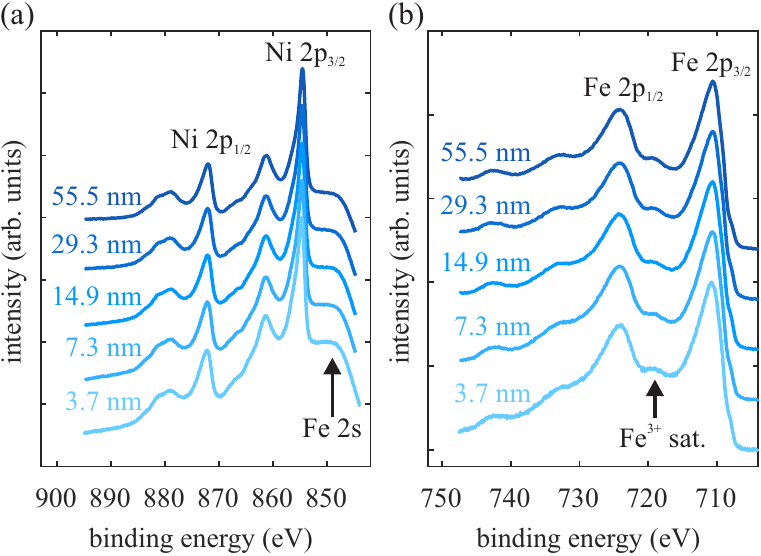}
	\caption{(a) Ni 2p and (b) Fe 2p core level HAXPES measurements of all prepared NFO films with varying film thickness.}
	\label{fig:HAXPES-Results}
\end{figure}

In addition to XPS, HAXPES measurements with an excitation energy of 6 keV were performed. 
Due to the higher kinetic energy of the photoelectrons in contrast to XPS, the probing depth of HAXPES is significantly higher. 
Thus, chemical composition and cationic valence states not only at the surface, but also in deeper layers of the NFO films can be determined. 
The respective Ni 2p and Fe 2p HAXPES spectra are depicted in Fig. \ref{fig:HAXPES-Results}. 
The recorded Ni 2p spectra in Fig. \ref{fig:HAXPES-Results} (a) show similar shapes and intensities for all film thicknesses with the spin-orbit split Ni 2p$_{1/2}$ and Ni 2p$_{3/2}$ peaks and the corresponding satellites $\sim$7 eV above their binding energies. 
Here, the superposition with the Fe 2s signal is stronger compared to XPS measurements due to the relatively higher photoelectric cross-section of the Fe 2s orbital at higher excitation energies. 
The shape of the spectra closely resembles the shape of the reference spectrum of NiO shown in Fig. \ref{fig:XPS_Ni2pFe2p} (a) with the exception of the high binding energy shoulder of the Ni 2p$_{3/2}$ in NiO. 
Due to the fact that this shoulder is characteristic for the presence of NiO clusters or NiO films thicker than one monolayer \cite{Veenendaal1993, Alders1996}, we can conclude that the prepared NFO films do not contain such NiO agglomerations. 
Nevertheless, the satellite structure in the spectra are similar to the NiO reference spectrum, which is consistent with the presence of Ni$^{2+}$ cations as it is also the case for NiO.

The respective Fe 2p core level spectra in Fig. \ref{fig:HAXPES-Results} (b) are also similar for all prepared films. 
Each spectrum exhibits a distinct satellite between the Fe 2p$_{1/2}$ and Fe 2p$_{3/2}$ signals at $\sim$718.9(2) eV, which is consistent with the satellite observed in Fe$_2$O$_3$ (see Fe$_2$O$_3$ reference spectrum in Fig. \ref{fig:XPS_Ni2pFe2p} (b) for comparison) and therefore characteristic for Fe$^{3+}$ valence states. 
In addition, the Fe 2p$_{1/2}$ and Fe 2p$_{3/2}$ binding energies with 724.3(2) eV and 710.8(2) eV, respectively, coincide with the positions of Fe$^{3+}$ in Fe$_2$O$_3$ reported in literature \cite{Yamashita2008}. 
In summary, Ni 2p and Fe 2p spectra measured by HAXPES do not exhibit any difference compared to the spectra measured by XPS, pointing to a uniform vertical distribution of Ni$^{2+}$ and Fe$^{3+}$ cations throughout the whole film.

Furthermore, a quantitative analysis of the HAXPES spectra is performed by applying equation (\ref{eq:IntRatioXPS}) for the Ni 2p and Fe 2p orbitals and their respective differential photoionization cross sections from Trzhaskovskaya \textit{et al.} \cite{Trzhaskovskaya2001}. 
Due to the increased overlap of the Fe 2s and Ni 2p signals for the applied photon energy of 6 keV, an unambiguous determination of the Shirley background for the whole Ni 2p spectrum is not possible. 
Thus, only the Ni 2p$_{1/2}$ signal, which is the least influenced by the Fe 2s spectrum, is taken into account in the quantitative analysis. 
The results are shown in Table \ref{tab:compositions} along with the results obtained by surface sensitive XPS and complementary results determined by bulk sensitive XRF spectroscopy. 
The Ni/(Ni+Fe) ratios determined by the three techniques are consistent with each other, considering an experimental error of $\pm$5\% for each technique. 
Thus, quantitative HAXPES analysis reveals that all prepared NFO films exhibit the stoichiometric Ni/Fe ratio not only at the surface, but also in deeper layers. This result is confirmed by XRF spectroscopy. 

\renewcommand{\arraystretch}{1.4}
\begingroup
\squeezetable
\begin{table}[]
	\begin{ruledtabular}
		\caption{Cationic Ni amount [Ni/(Ni+Fe) ratio] of the prepared NFO films with varying film thickness D determined by surface sensitive XPS and bulk sensitive HAXPES and XRF.}
		\label{tab:compositions}
		\begin{tabular}{cccc}
			D (nm)  & XPS & HAXPES & XRF \\
			\hline
			3.7 & 34 \% & 30 \% & 36 \% \\
			7.3 & 34 \% & 32 \% & 35 \% \\
			14.9 & 36 \% & 33 \% & 34 \% \\
			29.3 & 37 \% & 33 \% & 34 \% \\
			55.5 & 36 \% & 36 \% & 35 \%
		\end{tabular}
	\end{ruledtabular}
\end{table}
\endgroup
\renewcommand{\arraystretch}{1}

\subsection{XRR}
\label{subsec:XRR}
\begin{figure}[b]
	\centering
	\includegraphics[width=\columnwidth]{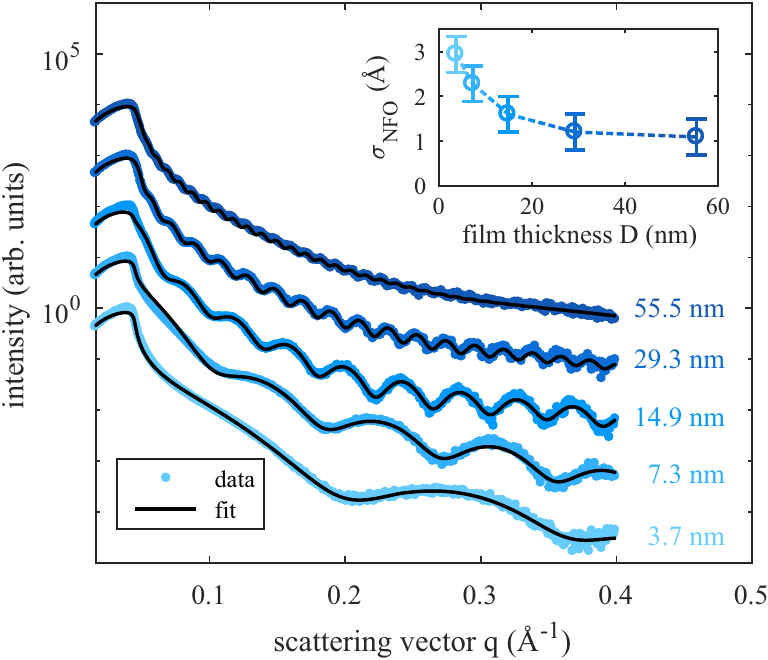}
	\caption{X-ray reflectivity measurements for all NFO films with varying thickness. The calculated XRR curves are in excellent agreement with the measured data. The NFO surface roughness $\sigma_\text{NFO}$ is decreasing for thicker films (shown in the inset).}
	\label{fig:XRR-Curves}
\end{figure}

Similar to HAXPES, x-ray reflectivity measurements of the NFO films were conducted after transport under ambient conditions. 
The results are depicted in Fig. \ref{fig:XRR-Curves}. 
For all film thicknesses, clear Kiessig-fringes are visible, which stem from interference of the beams reflected from the film surface and the interface between film and substrate. 
For the respective calculations, literature values of the refractive index of the STO substrate were used \cite{Henke1993}, allowing a deviation of $\pm$2\%. 
Film thicknesses, interface roughnesses and refractive index of the NFO films were used as free fit parameters. 
As seen in Fig. \ref{fig:XRR-Curves}, the calculated XRR curves are in excellent agreement with the corresponding measurements. 
The obtained dispersion $\delta_\text{NFO}^\text{meas}=1.63\times10^{-5}$ remains constant for all NFO films and is only $\sim$4\% higher than the literature value of $\delta_\text{NFO}^\text{lit}=1.57\times10^{-5}$ \cite{Henke1993}, also pointing to stoichiometric films.
The NFO surface roughness $\sigma_\text{NFO}$ decreases from 3.0 $\text{\AA}$ to 1.1 $\text{\AA}$ with increasing film thickness (see inset of Fig. \ref{fig:XRR-Curves}). 
This behavior is in accordance with the results obtained by LEED experiments (cf. section \ref{subsec:LEEDandXPS}), which are attributed to an increasing long-range order of the NFO surfaces with less defects for thicker films. 
In contrast, the NFO/STO interface roughness $\sigma_\text{STO}$ slightly rises only for the thickest NFO film, which results in a stronger damping of the Kiessig-fringes. 

\subsection{HR-XRD}
\begin{figure*}[htbp]
	\centering
	\includegraphics[width=\textwidth]{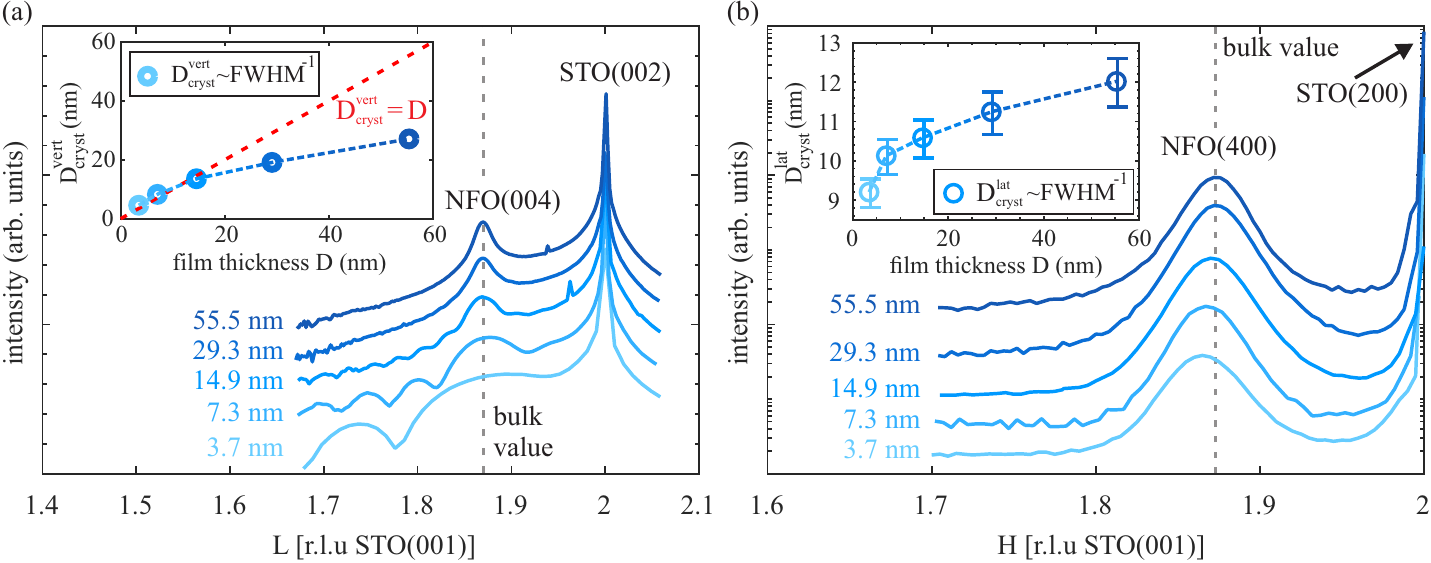}
	\caption{(a) X-ray diffraction mesaurements of the (00L) crystal truncation rod across the NFO(004) and STO(002) Bragg reflections. The vertical crystallite size D$_\text{cryst}^\text{vert}$ determined by the Scherrer formula (D$_\text{cryst}^\text{vert}$$\sim$FWHM(004)$^{-1}$) is clearly deviating from the film thickness to lower values for NFO films above 29 nm (shown in the inset). (b) Grazing incidence x-ray diffraction measurements along the (H00) direction across the NFO(400) and STO(200) reflexes. The FWHM of the NFO(400) reflex decreases meaning that the lateral crystallite size (D$_\text{cryst}^\text{lat}$$\sim$FWHM(400)$^{-1}$) increases for thicker films (shown in the inset).}
	\label{fig:004_400_results}
\end{figure*}
In order to characterize the structure of the prepared NFO films in detail, HR-XRD experiments were conducted at beamline P08 of DESY. 
The results of the XRD measurements along the out-of-plane (00L) and the in-plane (H00) directions are depicted in Fig. \ref{fig:004_400_results} (a) and (b), respectively. 

The (00L) scans were measured across the NFO(004) and the STO(002) Bragg reflections [cf. Fig. \ref{fig:004_400_results} (a)]. 
For film thicknesses up to 14.9 nm, the XRD scans exhibit clear Laue fringes resulting from a well-ordered NFO growth with smooth interfaces and homogeneous film thickness. 
These fringes vanish for the two thickest films, indicating lower ordering in vertical direction and larger interface roughness for thicker films. 
Furthermore, quantitative analysis is applied by peak fitting of the NFO(004) reflex. 
Its FWHM is used to estimate the vertical crystallite size D$_\text{cryst}^\text{vert}$ by the Scherrer equation \cite{Scherrer1918}. 
The vertical crystalline grain size in dependence of the film thickness is depicted in the inset of Fig. \ref{fig:004_400_results} (a). 
Here, films up to 14.9 nm thickness exhibit an almost completely single crystalline film in vertical direction, since the crystallite size coincides with the film thickness. 
For both thicker films, D$_\text{cryst}^\text{vert}$ clearly drops compared to the actual film thickness down to a value of $\sim$48 \% for the thickest film of 55.5 nm. 
This observation is in accordance with the decline of the Laue fringes for both thickest films as the formation of disordered layers also results in damping of the fringes. 

Furthermore, (H00) measurements across the NFO(400) and STO(200) reflexes, shown in Fig. \ref{fig:004_400_results} (b), reveal information about the lateral ordering of the NFO films. 
Similar to the NFO(004) reflection, the NFO(400) reflex was fitted by a Gaussian function to determine its FWHM and position for each film. 
For the FWHM, a steady decline for increasing film thickness is noted. 
By applying the Scherrer equation \cite{Scherrer1918}, equivalent to the analysis of the NFO(004) reflex, this behavior results in an increase of the lateral crystallite size D$_\text{cryst}^\text{lat}$ from $\sim$9 nm for the thinnest up to $\sim$12 nm for the thickest NFO film [shown in the inset of Fig. \ref{fig:004_400_results} (b)]. 
As described above, this behavior comes along with a decreasing vertical crystallite size compared to the total film thickness.

\begin{figure}[b]
	\includegraphics[width=\columnwidth]{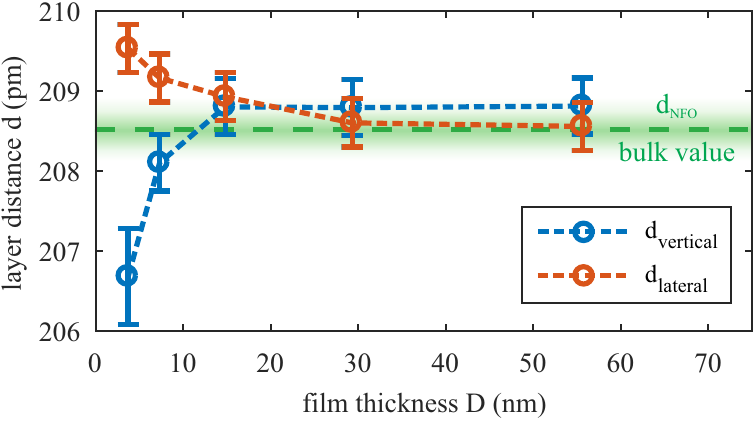}
	\caption{Vertical and lateral layer distances for varying NFO film thicknesses determined by the NFO(004) and NFO(400) positions, respectively. For films up to 7.3 nm thickness, an unexpected behavior of lateral tensile and vertical compressive strain is observed, followed by complete relaxation for thicker films.}
	\label{fig:latticeconstants}
\end{figure}

\begin{figure*}[t]
	\centering
	\includegraphics[width=\textwidth]{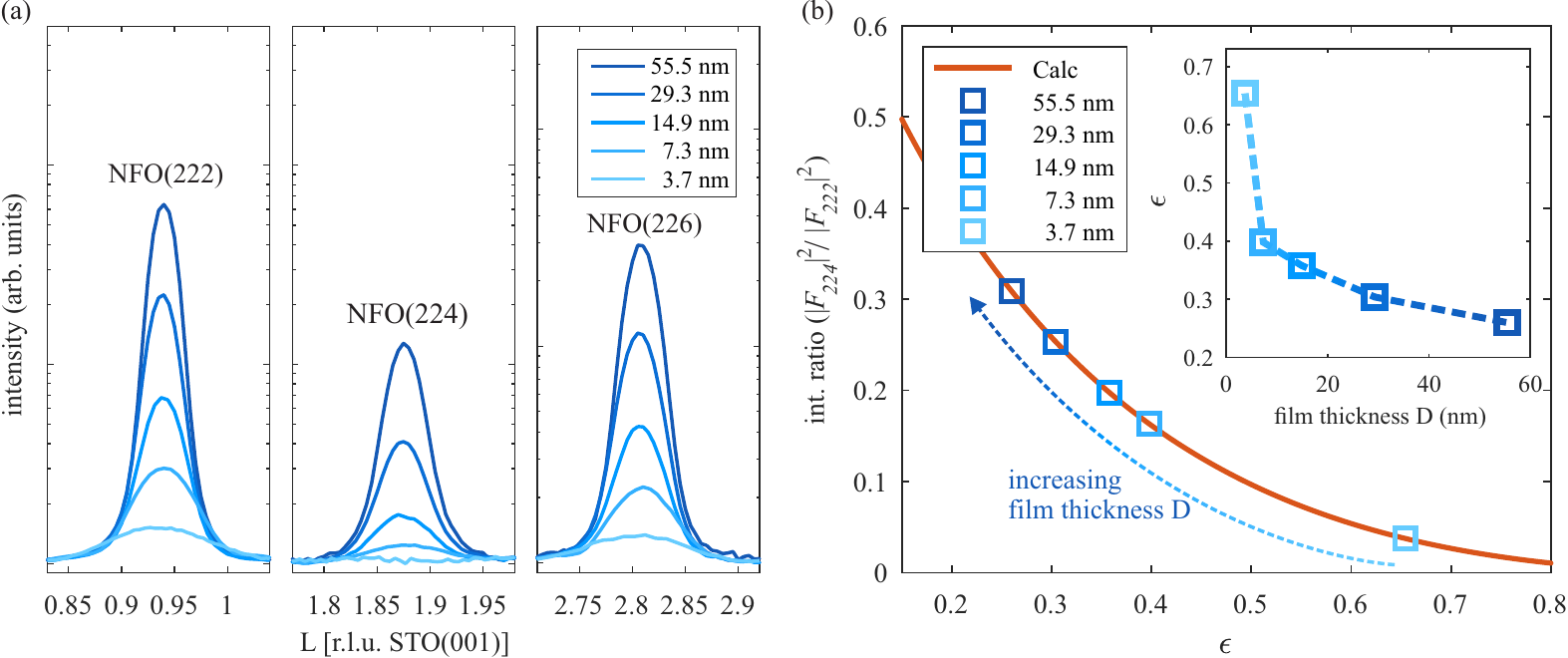}
	\caption{(a) GIXRD measurements along the NFO(22L) CTR across the NFO(222), NFO(224) and NFO(226) Bragg peaks. The increased intensity of the reflections for thicker films are due to the increased crystalline amount in the NFO films. Note the logarithmic scale on the intensity axis. (b) Calculated and experimental intensity ratios between the NFO(224) and NFO(222) reflexes as a function of the disorder parameter $\epsilon$. For increasing NFO film thicknesses, the occupancy of tetrahedral lattice sites and therefore the NFO(224)/NFO(222) intensity ratio increases, resulting in a decrease of $\epsilon$ (shown in the inset).}
	\label{fig:22L-Results}
\end{figure*}

Moreover, the vertical and lateral layer distances, derived from the NFO(004) and NFO(400) positions, respectively, are depicted in Fig. \ref{fig:latticeconstants} for varying film thicknesses. 
For thin films up to 7.3 nm, the vertical layer distance exhibits smaller values than for bulk NFO (d$_{\text{NFO}}$ = 208.5 pm). 
In contrast, the lateral layer distance for these film thicknesses deviates to higher values compared to the bulk value. 
This lateral tensile and vertical compressive strain contradicts an adaption of the film to the substrate lattice as the bulk NFO layer distance is 6.8\% larger than the STO(001) layer distance. 
Instead, following standard models of strain and relaxation in epitaxy, lateral compression and vertical tensile strain should be observed for an adapted NFO lattice to the STO substrate. 
Nevertheless, these observations are in accordance with previous studies on Fe$_3$O$_4$, NiFe$_2$O$_4$ or CoFe$_2$O$_4$ thin films on STO(001) \cite{Kuschel2017, Hoppe2015, Gao2009, Moyer2012}.

In addition, GIXRD scans along the NFO(22L) CTR were performed to study the cationic occupancy of tetrahedral ($T_d$) and octahedral ($O_h$) lattice sites, separately. 
The resulting measurements across the NFO(222), NFO(224) and NFO(226) Bragg peaks with varying film thickness are depicted in Fig. \ref{fig:22L-Results} (a). 
The rise in total intensities of the Bragg reflections for thicker films are due to the rising amount of crystalline NFO with increasing film thickness (cf. above). 
As the NFO(222) and NFO(226) reflections both originate from diffraction at cations on $O_h$ sites and O$^{2-}$ anions to the same extent, the difference in intensity between these two peaks for one film is due to the influence of the Debye-Waller factor. 
In contrast, only occupied $T_d$ sites contribute to the intensity of the NFO(224) peak. 
For a quantitative analysis, the Debye-Waller factor is determined by comparison of the NFO(222) and NFO(226) intensities. 
By this, the squared structure factor $\left|F_{HKL}\right|^2$ of each Bragg peak along the NFO(22L) CTR can be determined. 

Following the model of disorder described by Bertram \textit{et al.} \cite{Bertram2013}, the structure factor $F_{HKL}$ of the ferrite film can be described as a sum of the structure factor of nickel ferrite $F_{HKL}$(NFO) as an ideal inverse spinel and of a deficient rock salt like structure $F_{HKL}$(Ni$_{0.25}$Fe$_{0.5}$O). 
Here, Ni$_{0.25}$Fe$_{0.5}$O exhibits NFO stoichiometry, but a fraction of the vacant $O_h$ sites of the inverse spinel structure are occupied by cations taken from $T_d$ sites, while all $T_d$ sites are unoccupied. 
Therefore, the structure factor $F_{HKL}$ can be written as
\begin{equation}
\label{eq:F_HKL}
F_{HKL} = (1-\epsilon)F_{HKL}(\text{NFO}) + \epsilon F_{HKL}(\text{Ni}_{0.25}\text{Fe}_{0.5}\text{O}),
\end{equation}
where $\epsilon$ denotes the disorder parameter. 
For $\epsilon = 1$, only $O_h$ sites are occupied forming solely a deficient rock salt like structure, whereas for $\epsilon = 0$, $O_h$ and $T_d$ sites are occupied as in the ideal inverse spinel structure of NFO. 
Following this model, intensity ratios $I(224)/I(222) \propto \left|F_{224}\right|^2/\left|F_{222}\right|^2$ of the NFO(224) and NFO(222) peaks can be calculated by equation (\ref{eq:F_HKL}) and are depicted in Fig. \ref{fig:22L-Results} (b) as a function of $\epsilon$. 
For decreasing values of $\epsilon$, meaning increasing occupancy of tetrahedral lattice sites and therefore an increasing amount of NFO in the ideal inverse spinel structure, the intensity ratio $I(224)/I(222)$ increases. 
In contrast, increasing values for $\epsilon$, meaning higher probability of relocation from $T_d$ sites to $O_h$ vacancies, result in a decreasing intensity ratio between the NFO(224) and NFO(222) peaks, since the NFO(224) Bragg peak is forbidden, if cations occupy solely the $O_h$ sublattice. 
Experimental values for $\epsilon$ are determined by peak fitting of the NFO(224) and NFO(222) reflections and calculating their intensity ratios taking account of the Debye-Waller factor. 
The obtained results for all NFO films are also shown in Fig. \ref{fig:22L-Results} (b) in addition to the calculated data. 
A steady increase of the intensity ratio between the NFO(224) and NFO(222) peaks leads to decreasing disorder parameter values $\epsilon$ for increasing film thickness. 
Thus, thinner NFO films exhibit more deficient rock salt like structures, while the cationic occupancy of tetrahedral sites and thereby the amount of inverse spinel type structures increase for thicker films. 
In particular, for the thinnest film with a thickness of 3.7 nm a significant deviation of the disorder parameter $\epsilon$ compared to the thicker films is observed. 

\subsection{SQUID}
\label{subsec:SQUID}
\begin{figure}[b]
	\centering
	\includegraphics[width=1\columnwidth]{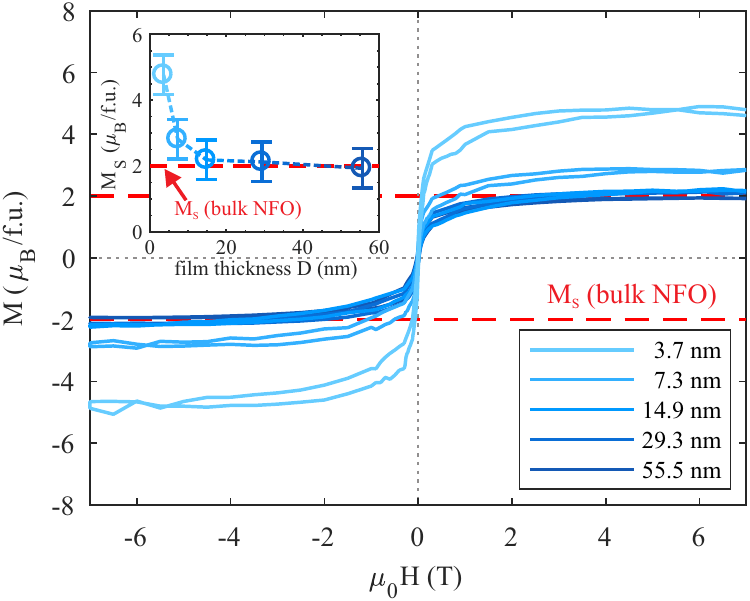}
	\caption{M vs. H curves for all NFO films with varying film thickness measured by SQUID magnetometry at 5 K after zero field cooling. The magnetic saturation is significantly enhanced for film thicknesses up to 7.3 nm. Thicker films exhibit the NFO bulk magnetization of 2 $\mu_B$/f.u. (shown in the inset).}
	\label{fig:SQUID-results}
\end{figure}

In addition to the chemical and structural characterization, magnetic properties of the NFO films were measured by SQUID magnetometry. 
Therefore, the sample was cooled down to 5 K in zero field to minimize thermal fluctuations during the measurements. 
The magnetization M was recorded in dependence of the applied magnetic field $\mu_0$H, which was tuned from +7 T to -7 T and back again in in-plane geometry parallel to the [100] direction of substrate and film.
In order to determine the separate magnetic response only from the NFO films, a linear background stemming from diamagnetic contributions from substrate and sample holder was subtracted from the data. 
The resulting magnetization was converted into units of $\mu_B$/f.u. (f.u.: formula unit). 
The resulting M vs. H curves for the NFO films with varying film thickness are shown in Fig. \ref{fig:SQUID-results}.

All films exhibit magnetic hysteresis loops characteristic for ferro-/ferrimagnetic (FM) material with very low coercive fields H$_\text{C}$ rising from only 10 mT for the thinnest up to 20 mT for the thickest NFO film. 
Magnetic saturation M$_\text{S}$ is reached for all films above a magnetic field of $\mu_0$H = 5 T. 
The magnetic saturations for films with thicknesses of 14.9 nm and above coincide with the value of 2 $\mu_B$/f.u. expected for bulk NFO, whereas this value is clearly exceeded for both thinner films. 
In particular, the 3.7 nm thin film exhibits a maximum M$_\text{S}$ of 4.8 $\mu_B$/f.u., far above the bulk value. 
This enhanced magnetization has already been observed for ultrathin NFO films on STO(001) prepared by RF sputtering or pulsed laser deposition \cite{Lueders2005, Hoppe2015}. 
In fact, the enhancement up to almost 250\% of the expected bulk magnetization is in agreement with the value obtained by Lüders \textit{et al.} for a 3 nm thin NFO film \cite{Lueders2005}. 

\section{Discussion}
Surface sensitive characterization by means of LEED and XPS, conducted \textit{in situ} directly after thin film deposition, reveals that all NFO films exhibit the expected surface structure and Ni:Fe ratios for stoichiometric NFO(001) grown on STO(001). 
In particular, quantitative analysis of XPS measurements results in an almost constant cationic Ni amount between 34\% and 37\%, agreeing with the amount of 33\% of stoichiometric NFO aimed for, considering an experimental error of $\pm$5\%. 
Furthermore, an excess of Fe$^{3+}$ compared to Fe$^{2+}$ is observed as expected for perfectly stoichiometric NFO, where only Fe$^{3+}$ cations should be present. 
Complementary HAXPES measurements also reveal a Ni/(Ni+Fe) ratio between 30\% and 36\%, proving that all films exhibit a uniform and stoichiometric Ni:Fe ratio not only at the surface, but also in bulk. 
XRF measurements confirm these results. 
Moreover, the observed Fe$^{3+}$ excess in surface sensitive XPS measurements is also clearly present in HAXPES, once more stating a uniform distribution of Ni and Fe cations in the stoichiometric ratio with the expected major Fe$^{3+}$ valence state throughout the whole film. 
The presence of other chemical phases like NiO, FeO, or metallic Ni or Fe are not observed in XPS or HAXPES and can therefore be excluded in all films.

In addition, LEED measurements show the expected square (1$\times$1) NFO surface structure already present for the thinnest film of 3.7 nm. 
Though, the intensity to background ratio is low and the FWHM of the diffraction spots is large compared to the flat and well-ordered substrate surface. 
This can be due to the incorporation of defects, e.g., misfit dislocations or vacancies, etc. in order to compensate for strain between film and substrate, which therefore results in a rougher surface with smaller lateral domain sizes. 
For increasing film thickness, spot intensities increase, whereas their FWHM continuously decrease, meaning higher long-range order with less defects. 
Thus, for thicker films a sufficient amount of defects were built in the film to compensate for strain and allowing relaxed film growth resulting in smoother surfaces with increasing lateral domain sizes at the surface. 
This assumption is corroborated by the results obtained by (GI)XRD measurements. 
Here, films with thicknesses up to 7.3 nm exhibit strained lattice constants in vertical and lateral direction. 
In contrast, thicker films with thicknesses of 14.9 nm and above exhibit complete relaxation. 
Moreover, the FWHM of the NFO(400) reflection decreases for increasing film thickness, meaning a growth of lateral grain sizes from $\sim$9 nm for the thinnest up to $\sim$12 nm for the thickest NFO film. 
In addition, the surface roughness of the NFO films, determined by XRR analysis, also exhibits a decay for thicker films, which is consistent with the observations made in GIXRD and LEED.

Analysis of the NFO(004) Bragg reflection reveals that the growth of lateral grains is accompanied by an increasing deviation of the vertical crystallite size from the film thickness for films above 14.9 nm thickness. 
For the thickest film of 55.5 nm, the most crucial deviation is noticed with a vertical crystallite size of only $\sim$48\% of the total film thickness. 
This determination is supported by the vanishing Laue fringes in the XRD scans of the respective NFO films, which is also attributed to either a rough substrate-film interface, disordered layers or poor crystallinity in vertical direction. 
Thus, the growth of lateral grains comes along with a decrease in growth of vertical crystallites. 
Both observations can still be explained by the progressive incorporation of strain releasing defects, e.g., dislocations, for increasing film thickness. 
These defects are likely to release strain in the film, resulting in bigger lateral grains, but also, by forming disordered layers, possibly separating vertical crytallites, which are in general not necessarily coherent with each other. 
Thus, the FWHM of the NFO(004) reflection as well as the Laue fringes are influenced by the diffraction of incoherently separated crystallites, resulting in a damping of the fringes and a bigger FWHM than expected for the total (summed up) crystallite size, which therefore deviates from the film thickness. 

More remarkably, the type of strain, which is present for film thicknesses up to 7.3 nm, is not expected for the growth of NFO(001) with a bigger (halved) lattice constant compared to STO(001). 
Then, by adaption of the NFO to the STO lattice, a lateral compression with a resulting vertical elongation of the NFO lattice constants should be expected. 
Here, the opposite behavior, i.e., lateral tensile and vertical compressive strain, is observed in the two thinnest NFO films. 
This unusual strain has already been reported in literature for the case of Fe$_3$O$_4$, NiFe$_2$O$_4$ or CoFe$_2$O$_4$ thin films on STO(001) \cite{Kuschel2017, Hoppe2015, Gao2009, Moyer2012}. 
Here, the assumption of auxetic behavior of ultrathin NFO films, proposed by Hoppe \textit{et al.} \cite{Hoppe2015}, can be excluded due to the distinct presence of tensile strain in lateral and compressive strain in vertical direction. 
This unexpected strain could further be due to a variety of complex mechanisms, including the influence of elastic, surface and interface energies \cite{Speck1994}, different thermal expansion factors of substrate and film \cite{Gao2009}, the influence of antiphase boundaries \cite{Arora2006}, etc.. 
However, a doubtless explanation for this behavior is still lacking and in need of further investigations.

In addition to strain analysis, the occupancy of lattice sites in the NFO films is determined by analyzing the NFO(22L) CTR. 
Here, a continuous rise of the NFO(224)/NFO(222) intensity ratio is observed for increasing film thickness. 
As the NFO(224) Bragg reflections solely originate from diffraction of occupied $T_d$ sites, this observation can be explained by an increase of the amount of spinel type structures. 
This behavior is analyzed by introducing the disorder parameter $\epsilon$, which is described as the probability for a cationic relocation from $T_d$ sites of the spinel type NFO to O$_h$ vacancies forming more defective rock salt type Ni$_{0.25}$Fe$_{0.5}$O structures with the same stoichiometry as inverse spinel NFO. 
Following this model, thin NFO films exhibit higher values for $\epsilon$ and, thus, higher amounts of rock salt like Ni$_{0.25}$Fe$_{0.5}$O, whereas the spinel type NFO amount rises for increasing film thickness. 
This tendency to lower tetrahedral occupancies for thin NFO films indicates a major presence of rock salt like structures close to the interface. 
This is in agreement with the observed formation of rock salt type FeO in the initial growth stages of thin Fe$_3$O$_4$ or $\alpha$-Fe$_2$O$_3$ films on different substrates as reported in literature \cite{Gota1999, Waddill2005, Schlueter2011, Bertram2012}. 
For the existence of rock salt type structures at the interface, two different models could be taken as a basis. 
As the determined stoichiometries are constant for all film thicknesses, there is no reason to assume a deviation from the NFO stoichiometry at the interface. 
Thus, in the first model, both rock salt type NiO and FeO structures could coexist at the interface with a majority of FeO to sustain the NFO stoichiometry. 
However, as XPS and HAXPES measurements do not exhibit any hint of either NiO or FeO (see sections \ref{subsec:LEEDandXPS} and \ref{subsec:HAXPES}), this model can be excluded directly. 
As a second more reasonable model, the already proposed deficient rock salt like Ni$_{0.25}$Fe$_{0.5}$O structure could form an ultrahin film on top of the substrate. 
As the NFO film thickness increases and the layer distances relax, more cations relocate from $O_h$ to $T_d$ sites, forming more inverse spinel like unit cells resulting in decreasing values for $\epsilon$. 
Here, it can be assumed that the strain present in the thinner films could have a major impact on the occupancy probability of cationic lattice sites. 
As the observed lateral tensile and vertical compressive strain also induce a change in the cation-oxygen bonding angles, it might be reasonable that the occupancy of $T_d$ sites becomes more unlikely. 
In contrast, as the film thickness increases, a relaxed fcc oxygen sublattice is formed with a higher probability of tetrahedral site occupancy. 

Furthermore, magnetic measurements show ferrimagnetic behavior of all NFO films, seen in the M vs. H hysteresis loops. 
As the three relaxed films for thicknesses of 14.9 nm and above exhibit magnetizations of $\sim$2 $\mu_B$/f.u. expected for bulk NFO, the two thinnest films exceed this value with a maximum of almost 250\% of the bulk value for the ultrathin 3.7 nm NFO film. 
This behavior, surprising at first sight, has already been reported for thin NFO films on STO \cite{Lueders2005, Hoppe2015}. 
Though, the explanations presented in these publications cannot withstand the observations made in this study. 
The ascription of the enhanced magnetization to an assumed auxetic behavior of ultrathin NFO films by Hoppe \textit{et al.} \cite{Hoppe2015} can be excluded, as the structural analysis, conducted here, reveals an amittedly unusual type of strain, but clearly no auxetic behavior. 
In contrast, Lüders \textit{et al.} \cite{Lueders2005} attributed the enhanced magnetization to a partial inversion of the NFO lattice from inverse to normal spinel. 
This assumption seems to be more likely, as the magnetic ordering of ferrites are predominantly influenced by the occupancy of the lattice sites in the spinel structure. 
Though, in this work, only a partial occupancy of the tetrahedral lattice sites is observed, as seen by GIXRD-analysis of the NFO(22L) crystal truncation rod. 
Thus, also a partial inversion, where Ni$^{2+}$ cations partly relocate to tetrahedral sites in exchange of Fe$^{3+}$ cations, cannot be the only explanation for the enhanced magnetization, as not all tetrahedral sites are occupied. 
Nevertheless, as our determination of site occupancies by GIXRD-analysis of the NFO(22L) CTR is not element sensitive, a partial inversion from inverse to normal spinel in the part of the spinel type NFO, as proposed by Lüders \textit{et al.}, could still contribute partly to the enhancement of the net magnetic moment.

In this study, the proposed deficient rock salt like Ni$_{0.25}$Fe$_{0.5}$O structure, predominantly existent for thinner films, could be one more likely reason for the observed magnetic enhancement.
Assuming that the exchange interactions are the same as in the spinel type structure (FM among $O_h$ sites, AFM between $T_d$ and $O_h$ sites), the complete cancellation of magnetic moments stemming from Fe$^{3+}$ cations on $T_d$ and $O_h$ sites is not fulfilled due to the deficient $T_d$ occupancy. 
Thus, the Fe$^{3+}$ cations on $O_h$ sites still contribute to the magnetization, leading to a higher net magnetic moment. 
However, perfect rock salt crystal structures solely exhibit AFM coupling between $O_h$ coordinated cations, which would lead to a lower net magnetic moment. 
Thus, the assumed FM exchange interaction between $O_h$ sites in the proposed deficient rock salt like Ni$_{0.25}$Fe$_{0.5}$O structures, as it is in the inverse spinel structure, is a hypothesis to be analyzed further. 
For instance, this problem may be attacked by X-ray Magnetic Circular Dichroism (XMCD), since this technique is element and site specific. 

Another reason for the enhanced magnetization could be found in the defective property of the rock salt like structures. 
Potzger \textit{et al.} observed unconventional ferromagnetism in defective pure ZnO powder after exposure to mechanical stress, although ZnO crystallizes in the rock salt structure and is non-magnetic. 
The observed ferromagnetism was related to defect creation in flakelike ZnO structures and was assumed to result from strain or domain boundaries induced by mechanical force \cite{Potzger2008}. 
This behavior could also appear in the proposed defective rock salt like Ni$_{0.25}$Fe$_{0.5}$O structures, as the strain induces smaller lateral grains with a higher lateral defect density for the thinnest NFO film.
Though, a distinct explanation for the defect induced ferromagnetism could not be given and is in need of further investigations.
 
In addition to this, a likely presence of oxygen vacancies, not detectable by XPS and HAXPES, would also alter the exchange interactions and could therefore cause an increase of the net magnetic moment.

\section{Summary}
In this study, NFO thin films with varying thickness were prepared on STO(001) by reactive molecular beam epitaxy. 
The chemical composition of the films at the surface and in bulk were characterized by means of XPS and HAXPES, respectively. 
Both techniques reveal a homogeneous cation distribution in the stoichiometric Ni:Fe ratio throughout the whole film with a majority of Ni$^{2+}$ and Fe$^{3+}$ valencies, as it is expected for ideal NFO. 
The presence of other chemical phases like NiO, FeO or metallic Ni or Fe can be excluded. 
A structural characterization of the film surface and the bulk structure was conducted using LEED, XRR and high resolution (GI)XRD. 
Thin NFO films exhibit a higher density of defects in lateral direction and at the surface than thicker films, resulting in an increase of lateral grain size with increasing film thickness. 
In contrast, vertical crystallites are assumed to be separated incoherently from each other due to the progressive incorporation of defects, most remarkable for thicker films. 
For thin films up to a thickness of 7.3 nm, an unexpected type of strain, i.e., vertical compressive and lateral tensile strain, is observed, which is in contradiction to an adaption of the NFO lattice to the substrate lattice. 
Nevertheless, thicker NFO films exhibit complete relaxation. 
This unusual strain is accompanied by a low occupancy probability of tetrahedral sites for thin films, resulting in large parts of deficient rock salt type Ni$_{0.25}$Fe$_{0.5}$O compared to spinel type NFO structures. 
These deficient rock salt type structures and the anomalous strain are assumed to be responsible for the significantly enhanced magnetization up to $\sim$250\% of the NFO bulk magnetization observed in ultrathin NFO films up to a thickness of 7.3 nm. 
For higher film thicknesses, the prepared NFO films exhibit the NFO bulk magnetization, confirming once more the NFO stoichiometry mainly present in the inverse spinel structure.

\begin{acknowledgments}
Financial support from the Deutsche Forschungsgemeinschaft (DFG under KU2321/6-1, and WO533/20-1) is gratefully acknowledged. 
We acknowledge DESY (Hamburg, Germany), a member of the Helmholtz Association HGF, for the provision of experimental facilities. 
Parts of this research were carried out at PETRA III and we would like to thank Andrei Gloskovskii for assistance in using beamline P22. 
Additionally, we gratefully acknowledge A. Becker, T. Peters, T. Kuschel and G. Reiss from Bielefeld University for providing measurement time at the x-ray diffractometer (XRR) and XRF spectrometer. 
\end{acknowledgments}

\bibliographystyle{apsrev4-1}
\bibliography{../literature/literature}

\begin{thebibliography}{33}%
\makeatletter
\providecommand \@ifxundefined [1]{%
 \@ifx{#1\undefined}
}%
\providecommand \@ifnum [1]{%
 \ifnum #1\expandafter \@firstoftwo
 \else \expandafter \@secondoftwo
 \fi
}%
\providecommand \@ifx [1]{%
 \ifx #1\expandafter \@firstoftwo
 \else \expandafter \@secondoftwo
 \fi
}%
\providecommand \natexlab [1]{#1}%
\providecommand \enquote  [1]{``#1''}%
\providecommand \bibnamefont  [1]{#1}%
\providecommand \bibfnamefont [1]{#1}%
\providecommand \citenamefont [1]{#1}%
\providecommand \href@noop [0]{\@secondoftwo}%
\providecommand \href [0]{\begingroup \@sanitize@url \@href}%
\providecommand \@href[1]{\@@startlink{#1}\@@href}%
\providecommand \@@href[1]{\endgroup#1\@@endlink}%
\providecommand \@sanitize@url [0]{\catcode `\\12\catcode `\$12\catcode
  `\&12\catcode `\#12\catcode `\^12\catcode `\_12\catcode `\%12\relax}%
\providecommand \@@startlink[1]{}%
\providecommand \@@endlink[0]{}%
\providecommand \url  [0]{\begingroup\@sanitize@url \@url }%
\providecommand \@url [1]{\endgroup\@href {#1}{\urlprefix }}%
\providecommand \urlprefix  [0]{URL }%
\providecommand \Eprint [0]{\href }%
\providecommand \doibase [0]{https://doi.org/}%
\providecommand \selectlanguage [0]{\@gobble}%
\providecommand \bibinfo  [0]{\@secondoftwo}%
\providecommand \bibfield  [0]{\@secondoftwo}%
\providecommand \translation [1]{[#1]}%
\providecommand \BibitemOpen [0]{}%
\providecommand \bibitemStop [0]{}%
\providecommand \bibitemNoStop [0]{.\EOS\space}%
\providecommand \EOS [0]{\spacefactor3000\relax}%
\providecommand \BibitemShut  [1]{\csname bibitem#1\endcsname}%
\let\auto@bib@innerbib\@empty
\bibitem [{\citenamefont {Hwang}\ \emph {et~al.}(2012)\citenamefont {Hwang},
  \citenamefont {Iwasa}, \citenamefont {Kawasaki}, \citenamefont {Keimer},
  \citenamefont {Nagaosa},\ and\ \citenamefont {Tokura}}]{Hwang2012}%
  \BibitemOpen
  \bibfield  {author} {\bibinfo {author} {\bibfnamefont {H.~Y.}\ \bibnamefont
  {Hwang}}, \bibinfo {author} {\bibfnamefont {Y.}~\bibnamefont {Iwasa}},
  \bibinfo {author} {\bibfnamefont {M.}~\bibnamefont {Kawasaki}}, \bibinfo
  {author} {\bibfnamefont {B.}~\bibnamefont {Keimer}}, \bibinfo {author}
  {\bibfnamefont {N.}~\bibnamefont {Nagaosa}},\ and\ \bibinfo {author}
  {\bibfnamefont {Y.}~\bibnamefont {Tokura}},\ }\href
  {https://doi.org/10.1038/nmat3223} {\bibfield  {journal} {\bibinfo  {journal}
  {Nature Materials}\ }\textbf {\bibinfo {volume} {11}},\ \bibinfo {pages}
  {103} (\bibinfo {year} {2012})}\BibitemShut {NoStop}%
\bibitem [{\citenamefont {Bauer}\ \emph {et~al.}(2012)\citenamefont {Bauer},
  \citenamefont {Saitoh},\ and\ \citenamefont {van Wees}}]{Bauer2012}%
  \BibitemOpen
  \bibfield  {author} {\bibinfo {author} {\bibfnamefont {G.~E.~W.}\
  \bibnamefont {Bauer}}, \bibinfo {author} {\bibfnamefont {E.}~\bibnamefont
  {Saitoh}},\ and\ \bibinfo {author} {\bibfnamefont {B.~J.}\ \bibnamefont {van
  Wees}},\ }\href {https://doi.org/10.1038/nmat3301} {\bibfield  {journal}
  {\bibinfo  {journal} {Nat. Mater.}\ }\textbf {\bibinfo {volume} {11}},\
  \bibinfo {pages} {391} (\bibinfo {year} {2012})}\BibitemShut {NoStop}%
\bibitem [{\citenamefont {Hoffmann}\ and\ \citenamefont
  {Bader}(2015)}]{Hoffmann2015}%
  \BibitemOpen
  \bibfield  {author} {\bibinfo {author} {\bibfnamefont {A.}~\bibnamefont
  {Hoffmann}}\ and\ \bibinfo {author} {\bibfnamefont {S.~D.}\ \bibnamefont
  {Bader}},\ }\href {https://doi.org/10.1103/PhysRevApplied.4.047001}
  {\bibfield  {journal} {\bibinfo  {journal} {Phys. Rev. Appl}\ }\textbf
  {\bibinfo {volume} {4}},\ \bibinfo {pages} {047001} (\bibinfo {year}
  {2015})}\BibitemShut {NoStop}%
\bibitem [{\citenamefont {Cibert}\ \emph {et~al.}(2005)\citenamefont {Cibert},
  \citenamefont {Bobo},\ and\ \citenamefont {L{\"u}ders}}]{Cibert2005}%
  \BibitemOpen
  \bibfield  {author} {\bibinfo {author} {\bibfnamefont {J.}~\bibnamefont
  {Cibert}}, \bibinfo {author} {\bibfnamefont {J.-F.}\ \bibnamefont {Bobo}},\
  and\ \bibinfo {author} {\bibfnamefont {U.}~\bibnamefont {L{\"u}ders}},\
  }\href {https://doi.org/10.1016/j.crhy.2005.10.008} {\bibfield  {journal}
  {\bibinfo  {journal} {C. R. Physique}\ }\textbf {\bibinfo {volume} {6}},\
  \bibinfo {pages} {977} (\bibinfo {year} {2005})}\BibitemShut {NoStop}%
\bibitem [{\citenamefont {Moussy}(2013)}]{Moussy2013}%
  \BibitemOpen
  \bibfield  {author} {\bibinfo {author} {\bibfnamefont {J.-B.}\ \bibnamefont
  {Moussy}},\ }\href {https://doi.org/10.1088/0022-3727/46/14/143001}
  {\bibfield  {journal} {\bibinfo  {journal} {J. Phys. D: Appl. Phys.}\
  }\textbf {\bibinfo {volume} {46}},\ \bibinfo {pages} {143001} (\bibinfo
  {year} {2013})}\BibitemShut {NoStop}%
\bibitem [{\citenamefont {Brabers}(1995)}]{Brabers1995}%
  \BibitemOpen
  \bibfield  {author} {\bibinfo {author} {\bibfnamefont {V.~A.~M.}\
  \bibnamefont {Brabers}},\ }\href@noop {} {\emph {\bibinfo {title} {{Handbook
  of Materials}}}},\ \bibinfo {edition} {1st}\ ed.,\ edited by\ \bibinfo
  {editor} {\bibfnamefont {K.~H.~J.}\ \bibnamefont {Buschow}},\ Vol.~\bibinfo
  {volume} {8}\ (\bibinfo  {publisher} {Elsevier},\ \bibinfo {year}
  {1995})\BibitemShut {NoStop}%
\bibitem [{\citenamefont {L{\"u}ders}\ \emph {et~al.}(2006)\citenamefont
  {L{\"u}ders}, \citenamefont {Barth\'{e}l\'{e}my}, \citenamefont {Bibes},
  \citenamefont {Bouzehouane}, \citenamefont {Fusil}, \citenamefont {Jacquet},
  \citenamefont {Contour}, \citenamefont {Bobo}, \citenamefont {Fontcuberta},\
  and\ \citenamefont {Fert}}]{Lueders2006}%
  \BibitemOpen
  \bibfield  {author} {\bibinfo {author} {\bibfnamefont {U.}~\bibnamefont
  {L{\"u}ders}}, \bibinfo {author} {\bibfnamefont {A.}~\bibnamefont
  {Barth\'{e}l\'{e}my}}, \bibinfo {author} {\bibfnamefont {M.}~\bibnamefont
  {Bibes}}, \bibinfo {author} {\bibfnamefont {K.}~\bibnamefont {Bouzehouane}},
  \bibinfo {author} {\bibfnamefont {S.}~\bibnamefont {Fusil}}, \bibinfo
  {author} {\bibfnamefont {E.}~\bibnamefont {Jacquet}}, \bibinfo {author}
  {\bibfnamefont {J.-P.}\ \bibnamefont {Contour}}, \bibinfo {author}
  {\bibfnamefont {J.-F.}\ \bibnamefont {Bobo}}, \bibinfo {author}
  {\bibfnamefont {J.}~\bibnamefont {Fontcuberta}},\ and\ \bibinfo {author}
  {\bibfnamefont {A.}~\bibnamefont {Fert}},\ }\href
  {https://doi.org/10.1002/adma.200500972} {\bibfield  {journal} {\bibinfo
  {journal} {Adv. Mater.}\ }\textbf {\bibinfo {volume} {18}},\ \bibinfo {pages}
  {1733} (\bibinfo {year} {2006})}\BibitemShut {NoStop}%
\bibitem [{\citenamefont {Matzen}\ \emph {et~al.}(2014)\citenamefont {Matzen},
  \citenamefont {Moussy}, \citenamefont {Wei}, \citenamefont {Gatel},
  \citenamefont {Cezar},\ and\ \citenamefont {Arrio}}]{Matzen2014}%
  \BibitemOpen
  \bibfield  {author} {\bibinfo {author} {\bibfnamefont {S.}~\bibnamefont
  {Matzen}}, \bibinfo {author} {\bibfnamefont {J.-B.}\ \bibnamefont {Moussy}},
  \bibinfo {author} {\bibfnamefont {P.}~\bibnamefont {Wei}}, \bibinfo {author}
  {\bibfnamefont {C.}~\bibnamefont {Gatel}}, \bibinfo {author} {\bibfnamefont
  {J.~C.}\ \bibnamefont {Cezar}},\ and\ \bibinfo {author} {\bibfnamefont
  {M.~A.}\ \bibnamefont {Arrio}},\ }\href {https://doi.org/10.1063/1.4871733}
  {\bibfield  {journal} {\bibinfo  {journal} {Appl. Phys. Lett.}\ }\textbf
  {\bibinfo {volume} {104}},\ \bibinfo {pages} {182404} (\bibinfo {year}
  {2014})}\BibitemShut {NoStop}%
\bibitem [{\citenamefont {Klewe}\ \emph {et~al.}(2014)\citenamefont {Klewe},
  \citenamefont {Meinert}, \citenamefont {Boehnke}, \citenamefont {Kuepper},
  \citenamefont {Arenholz}, \citenamefont {Gupta}, \citenamefont {Schmalhorst},
  \citenamefont {Kuschel},\ and\ \citenamefont {Reiss}}]{Klewe2014}%
  \BibitemOpen
  \bibfield  {author} {\bibinfo {author} {\bibfnamefont {C.}~\bibnamefont
  {Klewe}}, \bibinfo {author} {\bibfnamefont {M.}~\bibnamefont {Meinert}},
  \bibinfo {author} {\bibfnamefont {A.}~\bibnamefont {Boehnke}}, \bibinfo
  {author} {\bibfnamefont {K.}~\bibnamefont {Kuepper}}, \bibinfo {author}
  {\bibfnamefont {E.}~\bibnamefont {Arenholz}}, \bibinfo {author}
  {\bibfnamefont {A.}~\bibnamefont {Gupta}}, \bibinfo {author} {\bibfnamefont
  {J.-M.}\ \bibnamefont {Schmalhorst}}, \bibinfo {author} {\bibfnamefont
  {T.}~\bibnamefont {Kuschel}},\ and\ \bibinfo {author} {\bibfnamefont
  {G.}~\bibnamefont {Reiss}},\ }\href {https://doi.org/10.1063/1.4869400}
  {\bibfield  {journal} {\bibinfo  {journal} {J. Appl. Phys.}\ }\textbf
  {\bibinfo {volume} {115}},\ \bibinfo {pages} {123903} (\bibinfo {year}
  {2014})}\BibitemShut {NoStop}%
\bibitem [{\citenamefont {Hoppe}\ \emph {et~al.}(2015)\citenamefont {Hoppe},
  \citenamefont {D{\"o}ring}, \citenamefont {Gorgoi}, \citenamefont {Cramm},\
  and\ \citenamefont {M{\"u}ller}}]{Hoppe2015}%
  \BibitemOpen
  \bibfield  {author} {\bibinfo {author} {\bibfnamefont {M.}~\bibnamefont
  {Hoppe}}, \bibinfo {author} {\bibfnamefont {S.}~\bibnamefont {D{\"o}ring}},
  \bibinfo {author} {\bibfnamefont {M.}~\bibnamefont {Gorgoi}}, \bibinfo
  {author} {\bibfnamefont {S.}~\bibnamefont {Cramm}},\ and\ \bibinfo {author}
  {\bibfnamefont {M.}~\bibnamefont {M{\"u}ller}},\ }\href
  {https://doi.org/10.1103/PhysRevB.91.054418} {\bibfield  {journal} {\bibinfo
  {journal} {Phys. Rev. B}\ }\textbf {\bibinfo {volume} {91}},\ \bibinfo
  {pages} {054418} (\bibinfo {year} {2015})}\BibitemShut {NoStop}%
\bibitem [{\citenamefont {Seifikar}\ \emph {et~al.}(2012)\citenamefont
  {Seifikar}, \citenamefont {Calandro}, \citenamefont {Deeb}, \citenamefont
  {Sachet}, \citenamefont {Yang}, \citenamefont {Maria}, \citenamefont
  {Bassiri-Gharb},\ and\ \citenamefont {Schwartz}}]{Seifikar2012}%
  \BibitemOpen
  \bibfield  {author} {\bibinfo {author} {\bibfnamefont {S.}~\bibnamefont
  {Seifikar}}, \bibinfo {author} {\bibfnamefont {B.}~\bibnamefont {Calandro}},
  \bibinfo {author} {\bibfnamefont {E.}~\bibnamefont {Deeb}}, \bibinfo {author}
  {\bibfnamefont {E.}~\bibnamefont {Sachet}}, \bibinfo {author} {\bibfnamefont
  {J.}~\bibnamefont {Yang}}, \bibinfo {author} {\bibfnamefont {J.-P.}\
  \bibnamefont {Maria}}, \bibinfo {author} {\bibfnamefont {N.}~\bibnamefont
  {Bassiri-Gharb}},\ and\ \bibinfo {author} {\bibfnamefont {J.}~\bibnamefont
  {Schwartz}},\ }\href {https://doi.org/10.1063/1.4770366} {\bibfield
  {journal} {\bibinfo  {journal} {J. Appl. Phys.}\ }\textbf {\bibinfo {volume}
  {112}},\ \bibinfo {pages} {123910} (\bibinfo {year} {2012})}\BibitemShut
  {NoStop}%
\bibitem [{\citenamefont {Parratt}(1954)}]{Parratt1954}%
  \BibitemOpen
  \bibfield  {author} {\bibinfo {author} {\bibfnamefont {L.~G.}\ \bibnamefont
  {Parratt}},\ }\href {https://doi.org/10.1103/PhysRev.95.359} {\bibfield
  {journal} {\bibinfo  {journal} {Phys.~Rev.}\ }\textbf {\bibinfo {volume}
  {95}},\ \bibinfo {pages} {359} (\bibinfo {year} {1954})}\BibitemShut
  {NoStop}%
\bibitem [{\citenamefont {N\'{e}vot}\ and\ \citenamefont
  {Croce}(1980)}]{Nevot1980}%
  \BibitemOpen
  \bibfield  {author} {\bibinfo {author} {\bibfnamefont {L.}~\bibnamefont
  {N\'{e}vot}}\ and\ \bibinfo {author} {\bibfnamefont {P.}~\bibnamefont
  {Croce}},\ }\href {https://doi.org/10.1051/rphysap:01980001503076100}
  {\bibfield  {journal} {\bibinfo  {journal} {Revue de Physique Appliqu\'{e}e}\
  }\textbf {\bibinfo {volume} {15}},\ \bibinfo {pages} {761} (\bibinfo {year}
  {1980})}\BibitemShut {NoStop}%
\bibitem [{\citenamefont {Scofield}(1976)}]{Scofield1976}%
  \BibitemOpen
  \bibfield  {author} {\bibinfo {author} {\bibfnamefont {J.~H.}\ \bibnamefont
  {Scofield}},\ }\href
  {https://doi.org/https://doi.org/10.1016/0368-2048(76)80015-1} {\bibfield
  {journal} {\bibinfo  {journal} {Journal of Electron Spectroscopy and Related
  Phenomena}\ }\textbf {\bibinfo {volume} {8}},\ \bibinfo {pages} {129 }
  (\bibinfo {year} {1976})}\BibitemShut {NoStop}%
\bibitem [{\citenamefont {McIntyre}\ and\ \citenamefont
  {Cook}(1975)}]{McIntyre1975}%
  \BibitemOpen
  \bibfield  {author} {\bibinfo {author} {\bibfnamefont {N.~S.}\ \bibnamefont
  {McIntyre}}\ and\ \bibinfo {author} {\bibfnamefont {M.~G.}\ \bibnamefont
  {Cook}},\ }\href {https://doi.org/10.1021/ac60363a034} {\bibfield  {journal}
  {\bibinfo  {journal} {Anal. Chem.}\ }\textbf {\bibinfo {volume} {47}},\
  \bibinfo {pages} {2208} (\bibinfo {year} {1975})}\BibitemShut {NoStop}%
\bibitem [{\citenamefont {van Veenendaal}\ and\ \citenamefont
  {Sawatzky}(1993)}]{Veenendaal1993}%
  \BibitemOpen
  \bibfield  {author} {\bibinfo {author} {\bibfnamefont {M.~A.}\ \bibnamefont
  {van Veenendaal}}\ and\ \bibinfo {author} {\bibfnamefont {G.~A.}\
  \bibnamefont {Sawatzky}},\ }\href
  {https://doi.org/10.1103/PhysRevLett.70.2459} {\bibfield  {journal} {\bibinfo
   {journal} {Phys. Rev. Lett.}\ }\textbf {\bibinfo {volume} {70}},\ \bibinfo
  {pages} {2459} (\bibinfo {year} {1993})}\BibitemShut {NoStop}%
\bibitem [{\citenamefont {Alders}\ \emph {et~al.}(1996)\citenamefont {Alders},
  \citenamefont {Voogt}, \citenamefont {Hibma},\ and\ \citenamefont
  {Sawatzky}}]{Alders1996}%
  \BibitemOpen
  \bibfield  {author} {\bibinfo {author} {\bibfnamefont {D.}~\bibnamefont
  {Alders}}, \bibinfo {author} {\bibfnamefont {F.~C.}\ \bibnamefont {Voogt}},
  \bibinfo {author} {\bibfnamefont {T.}~\bibnamefont {Hibma}},\ and\ \bibinfo
  {author} {\bibfnamefont {G.~A.}\ \bibnamefont {Sawatzky}},\ }\href
  {https://doi.org/10.1103/PhysRevB.54.7716} {\bibfield  {journal} {\bibinfo
  {journal} {Phys. Rev. B}\ }\textbf {\bibinfo {volume} {54}},\ \bibinfo
  {pages} {7716} (\bibinfo {year} {1996})}\BibitemShut {NoStop}%
\bibitem [{\citenamefont {Yamashita}\ and\ \citenamefont
  {Hayes}(2008)}]{Yamashita2008}%
  \BibitemOpen
  \bibfield  {author} {\bibinfo {author} {\bibfnamefont {T.}~\bibnamefont
  {Yamashita}}\ and\ \bibinfo {author} {\bibfnamefont {P.}~\bibnamefont
  {Hayes}},\ }\href {https://doi.org/10.1016/j.apsusc.2007.09.063} {\bibfield
  {journal} {\bibinfo  {journal} {Appl.~Surf.~Sci.}\ }\textbf {\bibinfo
  {volume} {254}},\ \bibinfo {pages} {2441} (\bibinfo {year}
  {2008})}\BibitemShut {NoStop}%
\bibitem [{\citenamefont {Trzhaskovskaya}\ \emph {et~al.}(2001)\citenamefont
  {Trzhaskovskaya}, \citenamefont {Nefedov},\ and\ \citenamefont
  {Yarzhemsky}}]{Trzhaskovskaya2001}%
  \BibitemOpen
  \bibfield  {author} {\bibinfo {author} {\bibfnamefont {M.~B.}\ \bibnamefont
  {Trzhaskovskaya}}, \bibinfo {author} {\bibfnamefont {V.~I.}\ \bibnamefont
  {Nefedov}},\ and\ \bibinfo {author} {\bibfnamefont {V.~G.}\ \bibnamefont
  {Yarzhemsky}},\ }\href {https://doi.org/10.1006/adnd.2000.0849} {\bibfield
  {journal} {\bibinfo  {journal} {At.~Data Nucl.~Data Tables}\ }\textbf
  {\bibinfo {volume} {77}},\ \bibinfo {pages} {97} (\bibinfo {year}
  {2001})}\BibitemShut {NoStop}%
\bibitem [{\citenamefont {Henke}\ \emph {et~al.}(1993)\citenamefont {Henke},
  \citenamefont {Gullikson},\ and\ \citenamefont {Davis}}]{Henke1993}%
  \BibitemOpen
  \bibfield  {author} {\bibinfo {author} {\bibfnamefont {B.~L.}\ \bibnamefont
  {Henke}}, \bibinfo {author} {\bibfnamefont {E.~M.}\ \bibnamefont
  {Gullikson}},\ and\ \bibinfo {author} {\bibfnamefont {J.~C.}\ \bibnamefont
  {Davis}},\ }\href {https://doi.org/10.1006/adnd.1993.1013} {\bibfield
  {journal} {\bibinfo  {journal} {At. Data. Nucl. Data Tables}\ }\textbf
  {\bibinfo {volume} {54}},\ \bibinfo {pages} {181} (\bibinfo {year}
  {1993})}\BibitemShut {NoStop}%
\bibitem [{\citenamefont {Scherrer}(1918)}]{Scherrer1918}%
  \BibitemOpen
  \bibfield  {author} {\bibinfo {author} {\bibfnamefont {P.}~\bibnamefont
  {Scherrer}},\ }\href {http://eudml.org/doc/59018} {\bibfield  {journal}
  {\bibinfo  {journal} {Nachrichten von der Gesellschaft der Wissenschaften zu
  G{\"o}ttingen, Mathematisch-Physikalische Klasse}\ }\textbf {\bibinfo
  {volume} {1918}},\ \bibinfo {pages} {98} (\bibinfo {year}
  {1918})}\BibitemShut {NoStop}%
\bibitem [{\citenamefont {Kuschel}\ \emph {et~al.}(2017)\citenamefont
  {Kuschel}, \citenamefont {Spiess}, \citenamefont {Schemme}, \citenamefont
  {Rubio-Zuazo}, \citenamefont {Kuepper},\ and\ \citenamefont
  {Wollschl{\"a}ger}}]{Kuschel2017}%
  \BibitemOpen
  \bibfield  {author} {\bibinfo {author} {\bibfnamefont {O.}~\bibnamefont
  {Kuschel}}, \bibinfo {author} {\bibfnamefont {W.}~\bibnamefont {Spiess}},
  \bibinfo {author} {\bibfnamefont {T.}~\bibnamefont {Schemme}}, \bibinfo
  {author} {\bibfnamefont {J.}~\bibnamefont {Rubio-Zuazo}}, \bibinfo {author}
  {\bibfnamefont {K.}~\bibnamefont {Kuepper}},\ and\ \bibinfo {author}
  {\bibfnamefont {J.}~\bibnamefont {Wollschl{\"a}ger}},\ }\href
  {https://doi.org/10.1063/1.4995408} {\bibfield  {journal} {\bibinfo
  {journal} {Appl. Phys. Lett.}\ }\textbf {\bibinfo {volume} {111}},\ \bibinfo
  {pages} {041902} (\bibinfo {year} {2017})}\BibitemShut {NoStop}%
\bibitem [{\citenamefont {Gao}\ \emph {et~al.}(2009)\citenamefont {Gao},
  \citenamefont {Bao}, \citenamefont {Birajdar}, \citenamefont {Habisreuther},
  \citenamefont {Mattheis}, \citenamefont {Schubert}, \citenamefont {Alexe},\
  and\ \citenamefont {Hesse}}]{Gao2009}%
  \BibitemOpen
  \bibfield  {author} {\bibinfo {author} {\bibfnamefont {X.~S.}\ \bibnamefont
  {Gao}}, \bibinfo {author} {\bibfnamefont {D.~H.}\ \bibnamefont {Bao}},
  \bibinfo {author} {\bibfnamefont {B.}~\bibnamefont {Birajdar}}, \bibinfo
  {author} {\bibfnamefont {T.}~\bibnamefont {Habisreuther}}, \bibinfo {author}
  {\bibfnamefont {R.}~\bibnamefont {Mattheis}}, \bibinfo {author}
  {\bibfnamefont {M.~A.}\ \bibnamefont {Schubert}}, \bibinfo {author}
  {\bibfnamefont {M.}~\bibnamefont {Alexe}},\ and\ \bibinfo {author}
  {\bibfnamefont {D.}~\bibnamefont {Hesse}},\ }\href
  {https://doi.org/10.1088/0022-3727/42/17/175006} {\bibfield  {journal}
  {\bibinfo  {journal} {J. Phys. D: Appl. Phys.}\ }\textbf {\bibinfo {volume}
  {42}},\ \bibinfo {pages} {175006} (\bibinfo {year} {2009})}\BibitemShut
  {NoStop}%
\bibitem [{\citenamefont {Moyer}\ \emph {et~al.}(2012)\citenamefont {Moyer},
  \citenamefont {Kumah}, \citenamefont {Vaz}, \citenamefont {Arena},\ and\
  \citenamefont {Henrich}}]{Moyer2012}%
  \BibitemOpen
  \bibfield  {author} {\bibinfo {author} {\bibfnamefont {J.~A.}\ \bibnamefont
  {Moyer}}, \bibinfo {author} {\bibfnamefont {D.~P.}\ \bibnamefont {Kumah}},
  \bibinfo {author} {\bibfnamefont {C.~A.~F.}\ \bibnamefont {Vaz}}, \bibinfo
  {author} {\bibfnamefont {D.~A.}\ \bibnamefont {Arena}},\ and\ \bibinfo
  {author} {\bibfnamefont {V.~E.}\ \bibnamefont {Henrich}},\ }\href
  {https://doi.org/10.1063/1.4735233} {\bibfield  {journal} {\bibinfo
  {journal} {Appl. Phys. Lett.}\ }\textbf {\bibinfo {volume} {101}},\ \bibinfo
  {pages} {021907} (\bibinfo {year} {2012})}\BibitemShut {NoStop}%
\bibitem [{\citenamefont {Bertram}\ \emph {et~al.}(2013)\citenamefont
  {Bertram}, \citenamefont {Deiter}, \citenamefont {Schemme}, \citenamefont
  {Jentsch},\ and\ \citenamefont {Wollschläger}}]{Bertram2013}%
  \BibitemOpen
  \bibfield  {author} {\bibinfo {author} {\bibfnamefont {F.}~\bibnamefont
  {Bertram}}, \bibinfo {author} {\bibfnamefont {C.}~\bibnamefont {Deiter}},
  \bibinfo {author} {\bibfnamefont {T.}~\bibnamefont {Schemme}}, \bibinfo
  {author} {\bibfnamefont {S.}~\bibnamefont {Jentsch}},\ and\ \bibinfo {author}
  {\bibfnamefont {J.}~\bibnamefont {Wollschläger}},\ }\href
  {https://doi.org/10.1063/1.4803894} {\bibfield  {journal} {\bibinfo
  {journal} {J. Appl. Phys.}\ }\textbf {\bibinfo {volume} {113}},\ \bibinfo
  {pages} {184103} (\bibinfo {year} {2013})}\BibitemShut {NoStop}%
\bibitem [{\citenamefont {L\"uders}\ \emph {et~al.}(2005)\citenamefont
  {L\"uders}, \citenamefont {Bibes}, \citenamefont {Bobo}, \citenamefont
  {Cantoni}, \citenamefont {Bertacco},\ and\ \citenamefont
  {Fontcuberta}}]{Lueders2005}%
  \BibitemOpen
  \bibfield  {author} {\bibinfo {author} {\bibfnamefont {U.}~\bibnamefont
  {L\"uders}}, \bibinfo {author} {\bibfnamefont {M.}~\bibnamefont {Bibes}},
  \bibinfo {author} {\bibfnamefont {J.-F.}\ \bibnamefont {Bobo}}, \bibinfo
  {author} {\bibfnamefont {M.}~\bibnamefont {Cantoni}}, \bibinfo {author}
  {\bibfnamefont {R.}~\bibnamefont {Bertacco}},\ and\ \bibinfo {author}
  {\bibfnamefont {J.}~\bibnamefont {Fontcuberta}},\ }\href
  {https://doi.org/10.1103/PhysRevB.71.134419} {\bibfield  {journal} {\bibinfo
  {journal} {Phys. Rev. B}\ }\textbf {\bibinfo {volume} {71}},\ \bibinfo
  {pages} {134419} (\bibinfo {year} {2005})}\BibitemShut {NoStop}%
\bibitem [{\citenamefont {Speck}\ and\ \citenamefont
  {Pompe}(1994)}]{Speck1994}%
  \BibitemOpen
  \bibfield  {author} {\bibinfo {author} {\bibfnamefont {J.~S.}\ \bibnamefont
  {Speck}}\ and\ \bibinfo {author} {\bibfnamefont {W.}~\bibnamefont {Pompe}},\
  }\href {https://doi.org/10.1063/1.357097} {\bibfield  {journal} {\bibinfo
  {journal} {J. Appl. Phys.}\ }\textbf {\bibinfo {volume} {76}},\ \bibinfo
  {pages} {466} (\bibinfo {year} {1994})}\BibitemShut {NoStop}%
\bibitem [{\citenamefont {Arora}\ \emph {et~al.}(2006)\citenamefont {Arora},
  \citenamefont {Sofin}, \citenamefont {Shvets},\ and\ \citenamefont
  {Luysberg}}]{Arora2006}%
  \BibitemOpen
  \bibfield  {author} {\bibinfo {author} {\bibfnamefont {S.~K.}\ \bibnamefont
  {Arora}}, \bibinfo {author} {\bibfnamefont {R.~G.~S.}\ \bibnamefont {Sofin}},
  \bibinfo {author} {\bibfnamefont {I.~V.}\ \bibnamefont {Shvets}},\ and\
  \bibinfo {author} {\bibfnamefont {M.}~\bibnamefont {Luysberg}},\ }\href
  {https://doi.org/10.1063/1.2349468} {\bibfield  {journal} {\bibinfo
  {journal} {J. Appl. Phys.}\ }\textbf {\bibinfo {volume} {100}},\ \bibinfo
  {pages} {073908} (\bibinfo {year} {2006})}\BibitemShut {NoStop}%
\bibitem [{\citenamefont {Gota}\ \emph {et~al.}(1999)\citenamefont {Gota},
  \citenamefont {Guiot}, \citenamefont {Henriot},\ and\ \citenamefont
  {Gautier-Soyer}}]{Gota1999}%
  \BibitemOpen
  \bibfield  {author} {\bibinfo {author} {\bibfnamefont {S.}~\bibnamefont
  {Gota}}, \bibinfo {author} {\bibfnamefont {E.}~\bibnamefont {Guiot}},
  \bibinfo {author} {\bibfnamefont {M.}~\bibnamefont {Henriot}},\ and\ \bibinfo
  {author} {\bibfnamefont {M.}~\bibnamefont {Gautier-Soyer}},\ }\href
  {https://doi.org/10.1103/PhysRevB.60.14387} {\bibfield  {journal} {\bibinfo
  {journal} {Phys. Rev. B}\ }\textbf {\bibinfo {volume} {60}},\ \bibinfo
  {pages} {14387} (\bibinfo {year} {1999})}\BibitemShut {NoStop}%
\bibitem [{\citenamefont {Waddill}\ and\ \citenamefont
  {Ozturk}(2005)}]{Waddill2005}%
  \BibitemOpen
  \bibfield  {author} {\bibinfo {author} {\bibfnamefont {G.}~\bibnamefont
  {Waddill}}\ and\ \bibinfo {author} {\bibfnamefont {O.}~\bibnamefont
  {Ozturk}},\ }\href
  {https://doi.org/https://doi.org/10.1016/j.susc.2004.10.050} {\bibfield
  {journal} {\bibinfo  {journal} {Surface Science}\ }\textbf {\bibinfo {volume}
  {575}},\ \bibinfo {pages} {35 } (\bibinfo {year} {2005})}\BibitemShut
  {NoStop}%
\bibitem [{\citenamefont {Schlueter}\ \emph {et~al.}(2011)\citenamefont
  {Schlueter}, \citenamefont {Lübbe}, \citenamefont {Gigler},\ and\
  \citenamefont {Moritz}}]{Schlueter2011}%
  \BibitemOpen
  \bibfield  {author} {\bibinfo {author} {\bibfnamefont {C.}~\bibnamefont
  {Schlueter}}, \bibinfo {author} {\bibfnamefont {M.}~\bibnamefont {Lübbe}},
  \bibinfo {author} {\bibfnamefont {A.~M.}\ \bibnamefont {Gigler}},\ and\
  \bibinfo {author} {\bibfnamefont {W.}~\bibnamefont {Moritz}},\ }\href
  {https://doi.org/https://doi.org/10.1016/j.susc.2011.07.019} {\bibfield
  {journal} {\bibinfo  {journal} {Surface Science}\ }\textbf {\bibinfo {volume}
  {605}},\ \bibinfo {pages} {1986 } (\bibinfo {year} {2011})}\BibitemShut
  {NoStop}%
\bibitem [{\citenamefont {Bertram}\ \emph {et~al.}(2012)\citenamefont
  {Bertram}, \citenamefont {Deiter}, \citenamefont {Hoefert}, \citenamefont
  {Schemme}, \citenamefont {Timmer}, \citenamefont {Suendorf}, \citenamefont
  {Zimmermann},\ and\ \citenamefont {Wollschl{\"a}ger}}]{Bertram2012}%
  \BibitemOpen
  \bibfield  {author} {\bibinfo {author} {\bibfnamefont {F.}~\bibnamefont
  {Bertram}}, \bibinfo {author} {\bibfnamefont {C.}~\bibnamefont {Deiter}},
  \bibinfo {author} {\bibfnamefont {O.}~\bibnamefont {Hoefert}}, \bibinfo
  {author} {\bibfnamefont {T.}~\bibnamefont {Schemme}}, \bibinfo {author}
  {\bibfnamefont {F.}~\bibnamefont {Timmer}}, \bibinfo {author} {\bibfnamefont
  {M.}~\bibnamefont {Suendorf}}, \bibinfo {author} {\bibfnamefont
  {B.}~\bibnamefont {Zimmermann}},\ and\ \bibinfo {author} {\bibfnamefont
  {J.}~\bibnamefont {Wollschl{\"a}ger}},\ }\href
  {https://doi.org/10.1088/0022-3727/45/39/395302} {\bibfield  {journal}
  {\bibinfo  {journal} {J. Phys. D: Appl. Phys.}\ }\textbf {\bibinfo {volume}
  {45}},\ \bibinfo {pages} {395302} (\bibinfo {year} {2012})}\BibitemShut
  {NoStop}%
\bibitem [{\citenamefont {Potzger}\ \emph {et~al.}(2008)\citenamefont
  {Potzger}, \citenamefont {Zhou}, \citenamefont {Grenzer}, \citenamefont
  {Helm},\ and\ \citenamefont {Fassbender}}]{Potzger2008}%
  \BibitemOpen
  \bibfield  {author} {\bibinfo {author} {\bibfnamefont {K.}~\bibnamefont
  {Potzger}}, \bibinfo {author} {\bibfnamefont {S.}~\bibnamefont {Zhou}},
  \bibinfo {author} {\bibfnamefont {J.}~\bibnamefont {Grenzer}}, \bibinfo
  {author} {\bibfnamefont {M.}~\bibnamefont {Helm}},\ and\ \bibinfo {author}
  {\bibfnamefont {J.}~\bibnamefont {Fassbender}},\ }\href
  {https://doi.org/10.1063/1.2921782} {\bibfield  {journal} {\bibinfo
  {journal} {Applied Physics Letters}\ }\textbf {\bibinfo {volume} {92}},\
  \bibinfo {pages} {182504} (\bibinfo {year} {2008})},\ \Eprint
  {https://arxiv.org/abs/https://doi.org/10.1063/1.2921782}
  {https://doi.org/10.1063/1.2921782} \BibitemShut {NoStop}%
\end{thebibliography}%

\end{document}